\documentclass[11pt]{article}

\usepackage[latin1]{inputenc}  % para usar acentos
\usepackage{amsmath,amsfonts,amsthm,amscd,amssymb}  % These are to write mathematics
\usepackage{dsfont}			% Símbolos para os conjuntos R, N, etc.
\usepackage{thmtools}		% Teoremas

\usepackage[hyperindex,breaklinks]{hyperref}		% referências como links, \nameref e \autoref
\usepackage{geometry}		

\newtheorem{Corollary}{Corollary}

\theoremstyle{definition}
\newtheorem*{Definition}{Definition}

\theoremstyle{remark}
\newtheorem{remark}{Remark}

\declaretheoremstyle[
spaceabove=6pt, spacebelow=6pt,
headfont=\normalfont\bfseries,
notefont=\normalfont\bfseries, 
notebraces={}{},
bodyfont=\normalfont\itshape,
]{Estilo1}

\declaretheorem[style=Estilo1,numbered=no]{Axiom}
\declaretheorem[style=Estilo1,numbered=no,name=\!\!]{ThmName}

\newcommand\ket[1]			{\left| #1 \right\rangle}
\newcommand\braket[2] 		{\left\langle #1 \mid #2 \right\rangle}

\newcommand{\Schrodinger}			{Schr\"o\-din\-ger}
\newcommand{\HH}{\mathcal{H}}
\newcommand{\E}{\mathcal{E}}
\newcommand{\M}{\mathcal{M}}
\newcommand{\U}{\mathcal{U}}
\newcommand{\OO}{\mathcal{O}}
\newcommand{\R}{\mathcal{R}}

\begin{document}

\title{Analysis of Wallace's Proof of the Born Rule in Everettian Quantum Mechanics: Formal Aspects}

\author{Andr\'e L. G. Mandolesi \\ \emph{Departamento de Matem\'atica, Universidade Federal da Bahia} \\ \emph{Salvador-BA, Brazil} \\ \emph{E-mail:} \texttt{andre.mandolesi@ufba.br}}

\date{\today.}

\maketitle

\abstract{To solve the probability problem of the Many Worlds Interpretation of Quantum Mechanics, D.\,Wallace has presented a formal proof of the Born rule via decision theory, as proposed by D.\,Deutsch. The idea is to get subjective probabilities from rational decisions related to quantum measurements, showing the non-probabilistic parts of the quantum formalism, plus some rational constraints, ensure the squared modulus of quantum amplitudes play the role of such probabilities.

We provide a new presentation of  Wallace's proof, reorganized to simplify some arguments, and analyze it from a formal perspective. Similarities with classical decision theory are made explicit, to clarify its structure and main ideas. A simpler notation is used, and details are filled in, making it easier to follow and verify. Some problems have been identified, and we suggest possible corrections. 
}

%\tableofcontents

\section{Introduction}

This is the first of a set of articles analyzing the decision theoretic proof of the Born rule developed by D.\,Wallace. Here we focus on the formal aspects of deriving his result from the axioms. In other papers we discuss the more controversial aspects of his work, such as justifications for the axioms and interpretation of the result.

Despite its success, Quantum Mechanics still has unsolved conceptual problems, regarding measurements and the wavefunction collapse. Everettian Quantum Mechanics \cite{DeWitt1973,EverettIII1957}, or the Many Worlds Interpretation, tries to solve them by rejecting the Measurement Postulate, and applying the rest of the usual formalism to all systems, even macroscopic ones. In this theory, the quantum state of an observer, after a measurement, is a superposition of different versions of himself, each correlated to one of the results. The wavefunction collapse is illusory, due to the fact that each version is unaware of the others. Though this may seem far-fetched, it follows naturally from the quantum formalism minus the Measurement Postulate, and actually gives a clearer description of the measurement process than the usual Copenhagen Interpretation.

But as this theory solves old problems it creates new ones, such as the probability problem. It regards a possible disagreement with experiments, which seem to indicate that quantum measurements are probabilistic, with chances given by the Born rule. As this rule is removed with the Measurement Postulate, and the rest of the formalism is deterministic, it is not clear how to explain the observed probabilities. 

A possible solution is based on D.\,Deutsch's idea \cite{Deutsch1999} of using decision theory, with its well developed formalism of subjective probabilities, to derive the probabilistic part of quantum theory from the non-probabilistic one. He tried to show that, if a rational decision maker is to choose between bets on results of quantum measurements, quantum symmetries and rationality constraints imply he should decide as if results were probabilistic and followed the Born rule.

This proposal was met with criticism, leading to increasingly refined arguments, which culminated in D.\,Wallace's formal proof \cite{Wallace2010,Wallace2012}. Critics remain unconvinced, questioning some axioms or the meaning of his result, but there has been, so far, no criticism of the proof itself. And while supporters have proposed variations on Wallace's ideas, no attempts have been made to improve and clarify his formal presentation. 

Perhaps a reason for this is that his proof is hard to follow. Though he precedes it with an informal presentation, at times it seems to have little relation with the formal part. Also, a multilayered notation and terminology makes it hard to keep track of all formal details, while a number of typos, imprecisions and missing details can lead a reader astray. Combine that with the unfamiliar mix of quantum and decision theories, and the result is a maze of concepts and arguments which is arduous to navigate. This in turn makes the proof hard to verify, and the meaning of its result difficult to grasp.

In this article we give a new presentation of the proof, which, though following the same general lines as Wallace's, should be easier to analyze. Its structure was reorganized, allowing some demonstrations to be simplified, and highlighting its parallels with classical decision theory. Presented in the Appendix, it serves as an introduction to the ideas in a simpler context. 
We also identify some problems in Wallace's proof, suggesting corrections and improvements. 

In section \ref{sec:Preliminaries} we review the problems of the usual quantum theory, Everett's proposed solution, its probability problem, and the decision theoretic approach to solving it. 
Section \ref{sec:Quantum Decision Problem} presents Wallace's quantum decision problem and related concepts, as well as a glossary of his terminology. His axioms are stated in section \ref{sec:Wallace's Axioms}, which also includes an analysis of some of them and the concept of nullity. Section \ref{sec:Formal Proof} brings our presentation of his proof, with commentaries on its relations with the classical theory, its problems and possible corrections.   
Section \ref{sec:Conclusion} summarizes our conclusions.  In the Appendix we present classical decision theory, for those unfamiliar with it, as it helps in following Wallace's ideas.

\section{Preliminaries}\label{sec:Preliminaries}

We review some problems of the usual quantum theory, which motivated the Everettian alternative, and then present its main ideas. Its probability problem and some proposed solutions are discussed, in special the decision theoretic approach, on which Wallace's proof is based.

\subsection{The measurement problem and the quantum-classical transition}\label{sec:Measurement Problem}

In the usual Copenhagen interpretation of Quantum Mechanics (CQM), the Measurement Postulate states that if a system in a state 
\begin{equation}\label{eq:psi}
\ket{\psi}=\sum_i c_i\ket{i},
\end{equation}
with $\braket{i}{j} =\delta_{ij}$ and $\sum |c_i|^2=1$, is measured with respect to the basis $\{\ket{i}\}$, the result will be one (and only one) of the $i$'s, with the state collapsing to the corresponding $\ket{i}$. Also, results are probabilistic, according to the Born rule.

\begin{ThmName}[Born Rule]
The probability of result $i$ is given by
$p_i=w_i$,
where $w_i$ is its \emph{Born weight}, 
\begin{equation} \label{eq:Born_weight}
w_i=|c_i|^2=|\braket{i}{\psi}|^2.
\end{equation}
\end{ThmName}

This postulate agrees with experimental data, but is
conceptually ambiguous. Even though it sets measurements apart from other quantum processes, which obey the deterministic linear \Schrodinger\ equation, it lacks a precise definition of what are measurements. 
These might be distinguished for involving a classical macroscopic system, like an observer, but if this system's particles obey \Schrodinger's equation, how can they collectively produce a nonlinear probabilistic process? Which is random in principle, not simply due to a lack of knowledge about the states of the particles. And how does the collapse of the quantum state happen? 
Many attempts have been made to solve this \emph{measurement problem}, such as hidden variables theories, Bohmian mechanics, nonlinear \Schrodinger\ equations, and others, each with its own set of difficulties \cite{Auletta2000,Wheeler2014}.

This relates to the problem of whether Quantum Mechanics remains valid as systems get larger, with Classical Mechanics emerging from it. 
In the usual view, quantum superpositions should not happen at the macroscopic level, lest we observe \Schrodinger\ cats. But nothing in the quantum formalism seems to induce their disappearance in large systems, quite to the contrary. Some physicists consider Quantum Mechanics valid only for microscopic systems, with a new theory being needed to explain the quantum-classical transition. But this point of view becomes problematic as quantum phenomena are verified at increasingly larger scales, or for research in fields like quantum cosmology.
Some see decoherence \cite{Joos2003,Schlosshauer2007,Zurek2002} as a possible mechanism for the emergence of classicality. However, it is questionable whether it really eliminates superpositions, or just wipes out interference between their components, which remain nonetheless. 

\subsection{Everettian quantum mechanics}\label{sec:EQM}

H.\,Everett III's proposed solution \cite{EverettIII1957,DeWitt1973} is to eliminate the Measurement Postulate, and apply the rest of the quantum formalism even to  macroscopic systems. In this theory, evolution is deterministic at all times, following \Schrodinger's equation even during measurements.
It leads to macroscopic superpositions, but also explains why observers do not perceive them. If not for some unsolved problems, it might settle the problem of quantum measurements, and provide the missing link between quantum and classical mechanics. 

In Everettian Quantum Mechanics (EQM), a measurement is just entanglement of the measuring device with what is being measured. More precisely, a \emph{measuring device} for a basis $\{\ket{i}\}$ of a system is any apparatus, in a quantum state $\ket{D}$, interacting in such a way that, if the system is in state $\ket{i}$, the composite state evolves as\footnote{For simplicity, we assume the system remains in state $\ket{i}$, but this is not necessary.}
\begin{equation*} 
\ket{i}\otimes\ket{D}\ \  \longmapsto\ \  \ket{i}\otimes\ket{D_i},
\end{equation*}
where $\ket{D_i}$ is a new state of the device, registering result $i$.
Linearity of Schr\"odinger's equation implies that, if the system is in state \eqref{eq:psi}, the composite state evolves as
\begin{equation*} 
\ket{\psi}\otimes\ket{D} = \left(\sum_i c_i\ket{i}\right)\otimes\ket{D} \ \  \longmapsto\ \   \sum_i c_i\ket{i}\otimes\ket{D_i}.
\end{equation*}
This final state is to be accepted as an actual quantum superposition of macroscopic states. But it will not be perceived as such by an observer looking at the device, as, by the same argument, his state $\ket{O}$ will evolve into a superposition, according to
\begin{equation*}
\left(\sum_i c_i\ket{i}\otimes\ket{D_i}\right)\otimes\ket{O} \ \  \longmapsto\ \   \sum_i c_i\ket{i}\otimes\ket{D_i}\otimes\ket{O_i},
\end{equation*}
with $\ket{O_i}$ representing a state in which his brain registers seeing result $i$. 
By linearity, each component $\ket{i}\otimes\ket{D_i}\otimes\ket{O_i}$ evolves independently, as if the others did not exist, as long as interference is negligible. 
This condition is usually justified, for macroscopic systems, using decoherence arguments.

Everett's interpretation of this final state is that the observer has split into different versions of himself, each seeing a distinct result. Each version evolves as if the initial state had been $\ket{i}\otimes\ket{D}\otimes\ket{O}$, so he does not feel the splitting, nor the existence of his other versions.
Each component is called a \emph{world} or a \emph{branch}, and this evolution of one world into a superposition of many is called a \emph{branching process}. So in EQM all possible results of a measurement actually happen, but in different branches. The observer in state $\ket{O_i}$ only thinks the system has collapsed into $\ket{i}$ because he can not see the whole picture, with all other results and versions of himself.

Many problems plaguing CQM disappear in EQM, but new ones come along. The \emph{preferred basis problem} consists in how to decompose a macroscopic quantum state into branches behaving like the classical reality we observe, with negligible interference between them.
A solution has been proposed by Wallace \cite{Wallace2012}, using an adaptation of the decoherent histories formalism \cite{Gell-Mann1990,Gell-Mann1993} to EQM. The \emph{probability problem}, discussed in the next section, is how to make sense of quantum probabilities in EQM. 
Solving them would put EQM in a better theoretical standing than CQM, and might even reveal some testable difference between them.

\subsection{The probability problem}\label{sec:Probability_Problem}

In EQM, any result $i$ with $c_i \neq 0$ is obtained with certainty when measuring \eqref{eq:psi}, even if not all versions of the observer see it. The probability problem is reconciling this with experiments, which indicate results are probabilistic and follow the Born rule. 

It has a qualitative aspect, of how probabilities can emerge from a deterministic theory. In classical mechanics, processes can appear random due to our ignorance of details, but in EQM one must explain randomness even if the quantum state and its evolution are perfectly known.
Wallace \cite{Wallace2012} defends a purely operational and functional definition of probability, attained via decision theory and Bayesian inference. 
Other authors \cite{Vaidman1998,Saunders2010,Sebens2016} argue there is a \emph{self-locating uncertainty} in the time after the device measures the system, but before the observer sees the result. In their view, in this interval branching has already happened, but each version of the observer is still ignorant about his branch. 

There is also the quantitative aspect of accounting for probability values.
Everett \cite{EverettIII1957} proved that if a measure is attributed to branches, and is preserved by further branchings, it equals the Born weights \eqref{eq:Born_weight}. And, as the number of measurements tends to infinity, the total measure of  branches with results deviating from the Born rule tends to 0. For finite experiments, this means branches with frequencies deviating beyond a given error have small measure. But this only makes them negligible if Born weights have a probabilistic interpretation, leading to a circular argument. A similar idea was proposed by Graham \cite{Graham1973}, with the same problem.
Gleason's theorem \cite{Gleason1957} also implies the Born rule, if the probability of a branch does not depend on what other branches the decomposition basis has. 
But until we know how probabilities can emerge in EQM, we can not be sure they will satisfy the hypotheses of Everett or Gleason (of course, EQM is invalid if they do not).

Probabilities given by a counting measure, based on the idea that a measurement with $n$ results produces $n$ branches, might seem most natural for EQM. 
But they would disagree with Born's rule and quantum experiments: after many measurements, any sequence of results would appear in some branch, but in most branches the frequency of each result would tend to $1/n$, as if all results were equally probable.
Anyway, it might not be possible to count branches. As, in EQM, measurements are no different than other quantum processes, branchings can happen in all interactions, becoming a continuous and pervasive phenomenon. 
As most interactions involve few particles, many branches would be macroscopically similar, and a coarse-graining might reduce and stabilize their number. But it would be somewhat arbitrary, depending on the chosen fineness of grain.

Other attempts \cite{Albert1988,Buniy2006,Hanson2003,Zurek2005} have been made to explain why, in an \emph{Everettian universe} (i.e. one governed by EQM), quantum experiments would appear probabilistic, with probabilities given by the Born weights. 
We focus on Deutsch and Wallace's use of decision theory.

\subsubsection{The decision theoretic approach} \label{sec:decision theoretic approach}

 Decision theory (see \autoref{sec:Classical Decision Theory}) aims to explain how rational (in an axiomatically defined sense) decisions should be made. In cases of \emph{decision under risk}, in which there are multiple possible outcomes, with known probabilities, they follow the \nameref{th:PMEU}: choices with a higher expected value for the utilities of the outcomes are preferable.
When probabilities are unknown (\emph{decision under uncertainty}), Savage \cite{Savage1972} has shown that subjective probabilities can be obtained in a well defined way, and used to guide decisions via the same principle.

D.\,Deutsch \cite{Deutsch1999} proposed using this theory to obtain the Born rule from the non-probabilistic parts of the quantum formalism. The idea is that rational considerations and quantum symmetries, without any appeal to probabilities, should be enough to compel a decision maker to choose between bets, regarding the results of quantum experiments, using the following strategy:  
\begin{ThmName}[Born Strategy]\label{Born Strategy}
Decisions follow the \nameref{th:PMEU}, but with the expected utilities \eqref{eq:Expected Utility Classical} redefined as
\begin{equation}\label{eq:Expected Utility in Deutsch}
\mathrm{EU}=\sum_i w_i \cdot u(r_i),
\end{equation}
where the probabilities $p_i$ were replaced by the Born weights \eqref{eq:Born_weight} $w_i$ of the branches corresponding to each result of the bet.
\end{ThmName} 
In other words, he should behave in the same way classical decision makers do in probabilistic settings, except that Born weights play the role of probabilities.

Critics \cite{Baker2007,Barnum2000,Hemmo2007,Mallah2008,Price2006} contested Deutsch's proposal, expressing concerns about circularity, questioning his assumptions, proposing alternative decision strategies, or calling into question the meaning of his result. In response, there were many attempts \cite{Gill2005,Greaves2004,Saunders2004,Wallace2003,Wallace2007} to clarify his ideas, and finally a formal proof \cite{Wallace2010,Wallace2012} by D.\,Wallace.
Many authors \cite{Assis2011,Carroll.2014,Polley2001,Sebens2016,Wilson2012} consider this approach promising, and variations have been proposed in an effort to elucidate the situation. Others \cite{Albert2010,Finkelstein2009,Jansson2016,Kent2010,Maudlin2014,Price2010} remain unconvinced, questioning the ideas behind some of Wallace's axioms, presenting examples of  other 
reasonable decision strategies, or refusing to accept that there is any place for probability in EQM. 

But the debate has been focused on Wallace's informal ideas, and so far there has been no detailed criticism of his formal axioms and proof, which stand as the strongest defense of the decision theoretic approach. In this article we intend to fill such gap.

\section{Quantum Decision Problem}\label{sec:Quantum Decision Problem}

In Wallace's latest formalization \cite{Wallace2010,Wallace2012} of Deutsch's ideas, the formal presentation is preceded by an informal one.
As our focus here is a formal analysis of the proof, discussion of ideas behind concepts and axioms will be limited, and we refer to \cite{Wallace2012} for further information. The reader is invited to take a look at the Appendix first, as the classical case is simpler and helps grasping the ideas in the proof. 

Wallace defines a \emph{quantum decision problem} as being specified by:
\begin{itemize}
\item A separable Hilbert space $\HH$.
\item A collection $\E$ of closed\footnote{We added this condition, to avoid the possibility of $(E^\perp)^\perp\neq E$, for example.} subspaces of $\HH$, with $\HH\in\E$, forming a complete Boolean algebra under countable operations of conjunction $\wedge$, disjunction $\vee$, and complement $\perp$ (see section \ref{sec:Events} for definitions). A \emph{partition} of $E\in\E$ is defined as a set of mutually orthogonal elements of $\E$ whose disjunction is $E$.
\item A subset $\M\subset\E$ such that any $E\in\E$ has a partition in elements of $\M$.
\item A finite\footnote{We added this assumption for simplicity. See section \ref{sec:Rewards} for a discussion.} partition $\R$ of $\HH$.
\item For each $E\in\E$, a set $\U_E$ of unitary operators from $E$ into $\HH$.\footnote{Wallace includes some requirements, which we placed with the Richness Axioms (section \ref{sec:Richness Axioms}).}
\end{itemize}

We say the problem is \emph{rich} if all richness axioms (section \ref{sec:Richness Axioms}) are satisfied.

Wallace describes \cite[p.163]{Wallace2012} such problem as one in which, in an Everettian universe, a system prepared in some state is to be measured in a given basis. Bets are available, giving a payoff in each branch, depending on the result in it. And an Everettian agent (someone who knows EQM governs his Universe, and knows the Born weights of that state in the given basis) has to decide which bets he prefers. 

In this setting, $\HH$ is the Hilbert space of the total system of interest (including macroscopic elements, such as the agent, measuring device, payoffs, etc.). As the evolution of open quantum systems lacks unitarity, which is essential to the proof, $\HH$ must also include the environment.
In the next sections we briefly describe the meaning of the other symbols. A detailed analysis of the concepts will be left for another article. 

A \emph{solution} to a quantum decision problem is given by the assignment, for each state $\psi$ of each $M\in\M$, of a \emph{preference order} $\succ^\psi$ on the elements of $\U_M$. From it, the symbols $\prec^\psi$, $\sim^\psi$, $\succcurlyeq^\psi$,  and $\preccurlyeq^\psi$ are defined as usual.
To be acceptable, the solution must satisfy Wallace's preference axioms (section \ref{sec:Preference Axioms}), and in such case we call it a \emph{Wallacean solution}.
Wallace would probably prefer the term \emph{rational solution}, but we would rather avoid value-laden labels. 

The notation $\succ^\psi$ is ambiguous if $\psi$ is in more than one $M\in\M$, but there should be no problem in leaving $M$ implicit. 
Given $M,N\in\M$ and $U,V\in\U_N$, if $\psi\in M$ and $M\subset N$ the \nameref{ax:StaSup} axiom implies $U\succ^{\psi} V \Leftrightarrow U\lvert_M\succ^{\psi} V\lvert_M$.

\subsection{Events}\label{sec:Events}

An element of $\E$ is called an \emph{event}, and intuitively it is the subspace spanned by all states satisfying some proposition. For example, we could have an event $E$ spanned by all states in which a spin measurement resulted up and the agent received \$10. 

Operators $\wedge$, $\vee$, $\perp$ play in Quantum Logic \cite{Birkhoff1936,Engesser2009} roles similar to the connectives AND, OR, NOT of Classical Logic. A \emph{conjunction} $\wedge_i E_i$ is the intersection of subspaces $E_i$, a \emph{disjunction} $\vee_i E_i$ is the closure of the span of their union, and $E^\perp$ is the orthogonal complement of $E$. We also write $E\perp F$ meaning $E$ and $F$ are orthogonal, and $\dot{\vee}$ for disjunctions of orthogonal subspaces.

An important difference between quantum and classical (Boolean) logics is that the distributive law fails: if $S_u$, $S_d$ and $S_h$ are eigenspaces of spin up, down, and in some horizontal direction, then  $S_h\wedge(S_u\vee S_d) =S_h$ but $(S_h\wedge S_u)\vee(S_h\wedge S_d)=\{0\}$. 
So the requirement that $\E$ be a Boolean algebra\footnote{In \cite{Wallace2012}, Wallace gives, on pp. 152 and 435, good definitions of the Boolean condition. But on pp. 95 and 175 there are imprecise characterizations, which make it seem less restrictive than it actually is.} is quite strong, imposing a classical structure on it.
For any $E,F\in\E$, distributivity implies $E=(E\wedge F)\vee(E\wedge F^\perp)$, so the Boolean condition requires that events satisfy the following condition:
\begin{ThmName}[Orthogonality Condition]\label{ax:OrthCond}
The conjunction of two events $E,F\in\E$ is zero if, and only if, they are orthogonal, i.e.
$ E\wedge F=\{0\} \Leftrightarrow  E\perp F.$
\end{ThmName}
This implies the orthogonal projection of one event onto another is also an event, corresponding to their conjunction, i.e. $\Pi_F E=E\wedge F\in\E$. 

Given two partitions $\{E_i\}$ and $\{F_j\}$ of the same event, $\{F_j\}$ is a \emph{refinement} of $\{E_i\}$, and $\{E_i\}$ is a \emph{coarsening} of $\{F_j\}$, if each $E_i$ admits a partition in terms of $F_j$'s. With the \nameref{ax:OrthCond}, two partitions of an event always have a common refinement.

The requirement that $\E$ be complete is a technical condition, that every subset of the algebra has a supremum, necessary for operating with infinitely many elements. It is not really restrictive, as any Boolean algebra admits a unique completion.

\subsection{Macrostates}\label{sec:Macrostates}

Elements of $\M$ are called \emph{macrostates}. 
 Wallace says \cite[p.164]{Wallace2012} ``the choice of macrostates is largely fixed by decoherence, although the precise fineness of grain of the decomposition is underspecified''. It can not be too coarse, so ``an agent can be assumed not to care exactly what the microstate is within a given macrostate''. Also, ``an agent can have no practical control as to what state she gets, within a particular macrostate, on familiar statistical-mechanics and decoherence grounds'' \cite[p.170]{Wallace2012}. 

The intuitive idea is that a macrostate consists of macroscopically similar quantum states (how similar is up to coarsenings and refinements). It plays the role of a classical state, and should result from a solution to the preferred basis problem, which Wallace believes can be obtained via decoherence. For our formal analysis, it does not matter how $\M$ is formed, as long as the axioms of section \ref{sec:Wallace's Axioms} are satisfied.

Given a partition of $E\in\E$ into macrostates $M_i\in\M$, a state $\psi\in E$ has a \emph{branch decomposition} with \emph{branches} $\psi_i\in M_i$ if $\psi=\sum_i\psi_i$. If more than one $\psi_i$ is nonzero, we say $\psi$ is a \emph{branched state}.
The \nameref{ax:OrthCond} implies two branch decompositions of $\psi$ admit a common refinement.
As any $E\in\E$ has a partition in macrostates, intuitively events are disjunctions of macrostates satisfying some common condition, and $\psi\in E$ if it is decomposable in branches having such condition.

\subsection{Rewards}\label{sec:Rewards}

Elements of $\R$ are events called \emph{rewards}. They ``represent payoffs an agent could get'' \cite[p.175]{Wallace2012}, and are
``\ldots a coarse-graining of the macrostate subspaces\ldots such that an agent's only preference is to which reward subspace she is in'' \cite[p. 165]{Wallace2012}.

The description of $\R$ as a coarse-graining of $\M$ indicates that any $M\in\M$ should be in some $r\in\R$. 
But Wallace has not formalized such condition, and his example of $\M=\E$ \cite[p.176]{Wallace2012}, plus the use of $M\wedge r$ in the statement of \nameref{ax:MacIndif} \cite[p.179]{Wallace2012}, suggest otherwise. 
By the \nameref{ax:OrthCond}, $M=\dot{\vee}_{r\in\R} M\wedge r$, so a macrostate $M$ not contained in any $r$ is a disjunction of events from distinct reward subspaces. This seems to go against his characterizations of macrostates and rewards.

Our condition that $\R$ be finite is just to avoid some technicalities and focus on the main parts of Wallace's proof. As his goal is just to show Born weights replace probabilities in quantum decisions, and not to develop a general-purpose quantum decision theory, this is not an important loss of generality.
In any case, the proof can be adapted to work with infinitely many rewards, as in Wallace's original one.

\subsection{Acts}\label{sec:Acts}

Elements of $\U_E$ are \emph{acts available at $E\in\E$}. Intuitively, an act might represent the preparation of a quantum state, its measurement, placing a bet, receiving a payoff, or any other action of interest. 
In EQM, even macroscopic evolutions (of closed systems) are described by unitary operators. 
Availability of an act depends on $E$, e.g. the act of deciding a bet might only be available at events in which that bet has been placed.

Given $E\in\E$ and $U\in\U_E$, the range $U(E)$ might not be in $\E$, so Wallace uses $\OO_U$, the smallest event containing it (i.e. the intersection of all $F\in\E$ with $U(E)\subset F$). For $\psi\in E$, a partition  $\OO_U=\dot{\vee}_i M_i$, with $M_i\in\M$,  gives a branch decomposition of $U\psi$ in the $M_i$'s. The \nameref{ax:OrthCond} implies these are, up to coarsenings or refinements, the only possible branches resulting from the act.  

At a branched state there are many versions of the agent, each acting on his own branch $M_i$. In EQM their individual acts $U_i\in\U_{M_i}$ are restrictions of some $U\in\U_{\dot{\vee}_i M_i}$. In Wallace's terminology, they form a \emph{compatible act function}. Being the $M_i$'s mutually orthogonal, so must be their images $U_i(M_i)$.

\subsection{Notation and Glossary}\label{sec:Notation and Glossary}

A source of confusion in Wallace's text is an inconsistent notation. For example, he often uses $E$ to represent either an arbitrary event or a macrostate, at times without telling that it must be the latter.
To avoid this, we adopt the convention that:
\begin{itemize}
\item $\psi$, $\phi$, $\xi$ are always used for states (elements of $\HH$); 
\item $E$, $F$ for events (elements of $\E$);
\item $M$, $N$ for macrostates (elements of $\M$);
\item $r$, $s$, $t$ for rewards (elements of $\R$);
\item $U$, $V$, $W$, $X$ for acts (elements of some $\U_E$);
\item $\OO_U$, or $\OO(U)$, represents the smallest event containing the range of $U$;
\item $\Pi_E$ is the orthogonal projector onto $E$;
\item $U\lvert_F\in\U_F$ is the restriction of $U\in\U_E$ to $F\subset E$, i.e. $U\lvert_F\psi=U\psi$ for all $\psi\in F$;
\item $\mathds{1}_E$ is the identity map on $E$;
\item indices $i, j$ run over countable index sets.
\end{itemize}

Also, his terminology tends to make concepts and results seem simpler than they really are, while at the same time making it difficult to keep track of all formal details behind each term. In our presentation, we opted for a more explicit notation and terminology. To facilitate comparison, we provide a glossary of some of his terms:
\begin{itemize}
\setlength{\itemsep}{3pt}
\item an act $U$ is \emph{available} at an event $E$ if $U\in\U_E$;
\item an event $E$ is \emph{available} if $\U_E\neq\emptyset$;
\item a set of events $\{E_i\}$ is \emph{available} if $E_i\perp E_j$ for all $i\neq j$ and $\vee_i E_i$ is available; 
\item the \emph{weight} of an event $E$ with respect to a state $\psi$ and an act $U$ is 
\[\mathcal{W}_\psi(E|U) = \|\Pi_E U\psi\|^2/\|\psi\|^2;\]
\item a \emph{reward function} is a function $w:\R\rightarrow [0,1]$ such that $\sum\limits_{r\in\R} w(r)=1$;
\item the \emph{(characteristic) reward function} of $U$ and $\psi$ is
\[R_{\psi,U}(r)=\mathcal{W}_\psi(r|U) =\|\Pi_r U\psi\|^2/\|\psi\|^2;\]
\item $f[\alpha]$, $f_1[\alpha]$ and $f_2[\alpha]$ are certain reward functions depending on a parameter $\alpha$; 
\item an act \emph{has rewards in} $\mathcal{S}\subset\R$ if its range is a subset of $\vee \mathcal{S}$;
\item if $u$ is a real function on $\mathcal{S}\subset\R$ and an act $U$ has rewards in $\mathcal{S}$, the \emph{expected utility} of $U$ with respect to $\psi$ and $u$ is
\begin{equation*} 
\mathrm{EU}_\psi(U)=\sum_{r\in \mathcal{S}}R_{\psi,U}(r)\cdot u(r) =\sum_{r\in \mathcal{S}}\dfrac{\|\Pi_r U\psi\|^2}{\|\psi\|^2} \cdot u(r);
\end{equation*}
\item an \emph{utility function} is a real function $u$ on some $\mathcal{S}\subset\R$, given by his Utility Lemma; 
\item given a set $\mathcal{P}=\{p_i\}$ of positive numbers with $\sum_i p_i=1$, and a macrostate $M$ in a reward $r$, a  \emph{$\mathcal{P}$-branching} of $M$ is some $U\in\U_M$ such that $\OO_U\subset r$ and there is a partition $\OO_U=\vee_i M_i$ by macrostates with $\mathcal{W}_\psi(M_i|U)=p_i$ for any $\psi\in M$;
\item an \emph{erasure} of states $\psi,\psi'$ in macrostates $M,M'$ contained in the same reward $r$ is a pair of acts $U\in\U_M$, $U'\in\U_{M'}$ such that $\OO_U, \OO_{U'}\subset r$ and $U\psi=U'\psi'$;
\item an \emph{act function} $\mathcal{U}$ for an available set of events $\{E_i\}$ is a function assigning to each $E_i$ an act $\mathcal{U}(E_i)\in\U_{E_i}$; 
\item an act function $\mathcal{U}$ is \emph{compatible} if there is some $U\in\U_{\vee_iE_i}$ such that $U\lvert_{E_i}=\mathcal{U}(E_i)$;
\item a \emph{state dependent solution} to a decision problem is an assignment, for every available $M\in\M$ and every $\psi\in M$, of a two-place relation $\succ^\psi$ on the acts available at $M$;
\item an event $E$ is \emph{null} for a given state $\psi$ and act $U$ iff, whenever acts $V_1$ and $V_2$  are identical on the complement of $E$, $V_1U \sim^\psi V_2U$.
\end{itemize}

\section{Wallace's Axioms}\label{sec:Wallace's Axioms}

Except for a few corrections, and a diferent notation, we present Wallace's axioms as stated in \cite{Wallace2012}. Their meaning and justification will be only briefly discussed, as for a formal analysis of the proof these are irrelevant. We also do not question them here, leaving such line of inquiry for another paper.

The axioms are organized in two sets: richness axioms,  which are conditions on the sets $\U_E$ giving the agent a good selection of acts to consider, and preference axioms, conditions his  preference orders $\succ^\psi$ must satisfy. 

\subsection{Richness axioms}\label{sec:Richness Axioms}

We include here some conditions which Wallace placed in his definition of $\U_E$.

\begin{ThmName}[Restriction]\label{ax:Restr}
Let $E,F\in\E$ with $F\subset E$. If  $U\in\U_E$ then  $U\lvert_F\in\U_F$. 
\end{ThmName}

\begin{ThmName}[Composition]\label{ax:Compos}
Let $E\in\E$. If $U\in\U_E$ and $V\in\U_{\OO_U}$ then $VU\in\U_E$.
\end{ThmName}

\begin{ThmName}[Indolence]\label{ax:Indol}
Let $E\in\E$. If \ $\U_E\neq\emptyset$ then $\mathds{1}_E\in\U_E$.
\end{ThmName}

\begin{ThmName}[Continuation]\label{ax:Cont}
Let $E\in\E$ and $U\in\U_E$. Then $\U_{\OO_U}\neq\emptyset$.
\end{ThmName}

Events with $\U_E=\emptyset$ play no role in the problem, so it seems that Indolence and Continuation can be replaced by a simpler axiom, stating that $\mathds{1}_E\in\U_E$ for all $E\in\E$.

\begin{ThmName}[Irreversibility]\label{ax:Irrev}
Let $E,F\in\E$. If $E\perp F$ then $\OO_{U\lvert_E}\perp \OO_{U\lvert_F}$, for any $U\in\U_{E\dot{\vee} F}$.
\end{ThmName}

We changed Wallace's statement to include the hypothesis $E\perp F$, without which the axiom gives absurd results (e.g. if $E=F$). We also replaced $\OO_{U\lvert_E}\wedge \OO_{U\lvert_F}=\{0\}$ by $\OO_{U\lvert_E}\perp \OO_{U\lvert_F}$, which is equivalent (under the \nameref{ax:OrthCond}), seems more natural, and is what is needed for the proof.

By unitarity $U(E)\perp U(F)$, so the axiom requires that orthogonality be preserved when passing to the smallest events containing these ranges. 
With the \nameref{ax:OrthCond}, $\OO_{U\lvert_E}\perp \OO_{U\lvert_F}$ means states of $U(E)$ and $U(F)$ have no common branches, i.e. components in the same macrostate. 
So we have a \emph{branching structure}: as distinct branches evolve, they can not generate a common subbranch, and do not interfere.

\begin{ThmName}[Reward Availability]\label{ax:ReAv}
Given a set $\{M_i\}$ of mutually orthogonal macrostates\footnote{Wallace includes the condition $\U_{\vee_iM_i}\neq\emptyset$, which does not seem to play any role.}, and for each $i$ a reward $r_i\in\R$, there is some $U\in\U_{\vee_iM_i}$ such that $U(M_i)\subset r_i$ for all $i$.
\end{ThmName}

Wallace's justification for the availability of such \emph{reward acts} is that he is considering a ``relatively stylized decision problem'' and ``envelopes of cash can always be given to people'' \cite[p.167]{Wallace2012}.

\begin{ThmName}[Branching Availability]\label{ax:BrAv}
Let there be given:
\begin{itemize}
\item a set $\{M_i\}$ of mutually orthogonal macrostates, with each $M_i$ in some $r_i\in\R$;\footnote{Wallace omits this last condition, which is required by his definition of $\mathcal{P}$-branchings. As discussed in section \ref{sec:Rewards}, it is not clear if he assumes that every $M\in\M$ is in some $r\in\R$.}
\item for each $i$, a nonzero $\psi_i\in M_i$ and a set $\{p_{ij}\}$ of positive numbers with $\sum_j p_{ij}=1$.
\end{itemize}
Then there is $U\in\U_{\vee_iM_i}$ such that, for each $i$, 
\begin{itemize}
\item $U(M_i)\subset r_i$;
\item there is a partition $\OO_{U\lvert_{M_i}}=\dot{\vee}\!_j\, N_{ij}$ with $N_{ij}\in\M$ and  $\|\Pi_{N_{ij}} U\psi_i\|^2/\|\psi_i\|^2=p_{ij}$.
\end{itemize}
\end{ThmName}

Availability of such \emph{branching act} is explained by the possibility of preparing and measuring an arbitrary quantum state \cite[p.167]{Wallace2012}. 

We note here a small inconsistency in Wallace's concepts. He defines \cite[p.177]{Wallace2012} a $\mathcal{P}$-branching for a macrostate $M$, with $\mathcal{W}_\psi(M_i|U)=\|\Pi_{M_i} U\psi\|^2/\|\psi\|^2=p_i$ holding for all $\psi\in M$. But in his axiom he talks about $\mathcal{P}$-branchings of states. In our statement we require the condition to hold only for the $\psi_i$'s, which is enough for the proof. 

\begin{ThmName}[Erasure]\label{ax:Eras}
Let there be given:
\begin{itemize}
\item two sets $\{M_i\}$ and $\{N_i\}$ of macrostates, such that those in each set are mutually orthogonal, and, for each $i$, we have $M_i, N_i \subset r_i$ for some $r_i\in\R$;
\item for each $i$, nonzero states $\psi_i\in M_i$ and $\phi_i\in N_i$ with $\|\psi_i\|=\|\phi_i\|$.\footnote{Wallace omits this last condition, which is needed as his state vectors are not normalized \cite[p.176]{Wallace2012}.}
\end{itemize} 
Then there are $U\in\U_{\dot{\vee}_i M_i}$ and $V\in\U_{\dot{\vee}_i N_i}$ such that, for each $i$, $U(M_i), V(N_i)\subset r_i$ and $U\psi_i =V\phi_i$.
\end{ThmName}
To explain such \emph{erasures}, Wallace says \cite[p.167]{Wallace2012} the axiom ``effectively guarantees that an agent can just forget any facts about his situation that don't concern things he cares about (i.e. by definition: that don't concern where in the reward
space he is)''.
He describes $U$ and $V$ as taking $\psi_i$ and $\phi_i$ into an \emph{erasure subspace} of $r_i$, ``whose states correspond to the agent throwing the preparation system away after receiving the payoff but without recording the actual result of the measurement''. 
To justify why $U\psi_i$ and $V\phi_i$ are equal, he says that, as the agent ``lacks the fine control to know which act he is performing, all erasures should be counted as available if any are. It follows that, since for any two such agents all erasures are available, in particular there will be two erasures available satisfying the axiom'' \cite[p.167]{Wallace2012}.

\begin{ThmName}[Problem Continuity]\label{ax:PrCont}
Let $E\in\E$. Then $\U_E$ is an open subset of the set of unitary operators from $E$ to $\HH$, in the operator norm topology.
\end{ThmName}

Thus, if $U$ is available, so are all acts sufficiently close to it. 
The justification is that the agent can not control every microscopic detail of an act.

\subsection{Preference axioms}\label{sec:Preference Axioms}

Wallace calls these \emph{rationality axioms}, but we prefer a more neutral label. In another paper we will discuss whether they can be seen, like their counterparts in the classical theory, as mandates of rationality for decision problems.

\begin{ThmName}[Ordering]\label{ax:Ord}
For each $M\in\M$ and $\psi\in M$, $\succ^\psi$ is a total order on $\U_M$.
\end{ThmName}

This corresponds to \nameref{ax:Compl} and \nameref{ax:Trans} from classical decision theory. 

\begin{ThmName}[Branching Indifference]\label{ax:BrIndif}
Let $r\in\R$, $M\in\M$ with $M\subset r$, $\psi\in M$, and $U\in\U_M$.
If $U\psi\in r$ then $U\sim^\psi \mathds{1}_M$.
\end{ThmName}

Wallace justifies it by saying that 
``an agent doesn't care about branching per se: if a certain operation leaves his future selves in $N$ different macrostates but doesn't change any of their rewards, he is indifferent as to whether or not the operation is performed'', and that
``a  preference order which is not indifferent to branching per se would in practice be impossible to act on: branching is uncontrollable and ever-present in an Everettian universe'' \cite[p.170]{Wallace2012}.

Note that, despite the name and justification, the statement of the axiom makes no reference to branching acts. The only condition on $U$ is that it keeps $\psi$ in the same $r$. The axiom is used with both branching acts and erasures, in the \nameref{lm:Equivalence Lemma}. 

\begin{ThmName}[State Supervenience]\label{ax:StaSup}
Let $M, M'\in\M$, $\psi\in M$, $\psi'\in M'$, $U, V\in\U_{M}$ and $U', V'\in\U_{M'}$. If $U\psi=U'\psi'$ and $V\psi=V'\psi'$  then $U\succ^{\psi} V \Leftrightarrow U'\succ^{\psi'} V'$.
\end{ThmName}

So preferences can not depend on the acts or initial states, only on the final ones. 
In particular, $U\psi=V\psi$ implies $U\sim^\psi V$.

\begin{ThmName}[Solution Continuity]\label{ax:SolCont}
Let $M\in\M$, $\psi\in M$ and $U,V\in\U_M$. If $U\succ^\psi V$ then $U'\succ^\psi V'$ for any $U',V'\in\U_M$ sufficiently close (in the operator norm) to $U$ and $V$.
\end{ThmName}

The idea is that small perturbations of acts can not alter preferences, as agents can not distinguish arbitrarily similar acts, nor execute them with microscopic precision. 
It is reminiscent of the classical \nameref{ax:Arch} (Appendix \ref{sec:Decision under risk}).

\vspace{6pt}

Wallace includes two other preference axioms, \nameref{ax:MacIndif} and \nameref{ax:DiacCons}, and we have added one more, \nameref{ax:ActNDeg}. These will be discussed in more detail in the next sections. 

\subsubsection{Macrostate Indifference}\label{sec:Macrostate Indifference}

Wallace's initial description of this axiom is that ``an agent doesn't care what the microstate is provided it's within a particular macrostate'' \cite[p.170]{Wallace2012}. He justifies it saying that ``an agent can have no practical control as to what state she gets, within a particular macrostate, on familiar statistical-mechanics and decoherence grounds, and that we are interested in an agent's preferences only insofar as they show up in her actual dispositions to action''. So the idea is that decisions can not depend on microscopic details, for the agent has no such fine control, and if two states are different enough to affect his preferences, they should be in distinct macrostates.
In \cite[p.238]{Wallace2010} he even called the axiom Microstate Indifference, which better expresses the proposed idea.

Although his informal description makes no reference to rewards, they strangely appear in his formal statement \cite[p.179]{Wallace2012} (in expressions of the form $M\wedge r$, which seem odd in light of our discussion in section \ref{sec:Rewards}):

\begin{ThmName}[Macrostate Indifference]\label{ax:MacIndif}
Let $M,M', M_1,M_2\in\M$\footnote{Wallace does not say $M,M'\in\M$, but they must, as his preference order is only defined at macrostates.}, $\psi\in M$, $\psi'\in M'$, $U, V\in\U_{M}$, $U', V'\in\U_{M'}$, and $r_1,r_2\in\R$. If $\OO_{U},\OO_{U'}\subset M_1\wedge r_1$ and $\OO_{V},\OO_{V'}\subset M_2\wedge r_2$, then $U\succcurlyeq^{\psi} V \Leftrightarrow U'\succcurlyeq^{\psi'} V'$.
\end{ThmName}

If $M=M'$, $U=U'$ and $V=V'$, we get $U\succcurlyeq^{\psi} V \Leftrightarrow U\succcurlyeq^{\psi'} V$, so preferences do not depend on the initial microstates inside $M$ (given the other hypotheses). If $M=M'$ and $\psi=\psi'$, we find that different final microstates, in the same subspace of the form $M\wedge r$, do not matter either. This seems to agree with the informal description.

But the formal statement is much stronger. Even if $\psi$ and $\psi'$ are in different macrostates $M$ and $M'$, all that matters for the preferences are the subspaces $M_i\wedge r_i$ to which they are sent. Acts do not matter either, as with $\psi=\psi'$, $M_1=M_2$, $r_1=r_2$, $U=V'$ and $V=U'$ we get that all acts sending $\psi$ into $M_1\wedge r_1$ are equally preferred. Ultimately, preference between acts, with images in subspaces of the form $M\wedge r$ ($M\in\M, r\in\R$), is to depend only on such subspaces. 

Wallace says \cite[p.180]{Wallace2012} the axiom is only used\footnote{He also mentions it could be used to obtain Branching Indifference, if $\M=\E$.}, with \nameref{ax:ReAv} and \nameref{ax:BrIndif}, to prove the \nameref{def:OrdRe} (see section \ref{sec:Formal Proof}) is total. But he does not explain how, and it seems that extending the equivalence of preferences from $M\wedge r$ to the whole $r$ would also require \nameref{ax:StaSup}. And it might not even be enough, since not all states of $r$ must be connected by an available $U$. 

Perhaps his idea can be understood by noting that, in his presentation of classic decision theory \cite[p.458]{Wallace2012}, there is another axiom he also calls Macrostate Indifference. But that one, despite the name, makes no reference to macrostates, only to rewards. It in fact corresponds to replacing $M_i\wedge r_i$ in the above statement by just $r_i$, in which case the \nameref{def:OrdRe} would easily follow.

So it seems that there is a mix up of three different principles under the same name: an informal one, justified using the concept of macrostate; another, in terms of rewards, which was not clearly stated but could be used in the proof; and the formal one, which mixes both but turns out to be neither properly justified nor useful.

Anyway, none of them are really necessary, as we got the \nameref{def:OrdRe} from the \nameref{lm:Equivalence Lemma} (section \ref{sec:Formal Proof}), which relies only on the other axioms, and did not use \nameref{ax:MacIndif} for anything else.

\subsubsection{Nullity}\label{sec:Nullity}

The next axiom requires the concept of nullity.
Wallace calls \cite[p.178]{Wallace2012} an event $E$ \emph{null} for a state $\psi$ and an act $U$ iff, whenever acts $V_1$ and $V_2$  are identical on the complement of $E$, $V_1U \sim^\psi V_2U$. The idea is that the agent at $\psi$ does not care about what happens to his future (i.e. after $U$) selves (if any) at $E$. 
Wallace intends to prove this only happens because $\Pi_E U\psi=0$, i.e. the agent has no future selves at $E$.

This may seem like a straitforward adaptation of a similar concept from classical decision theory (see Appendix \ref{sec:Decision under uncertainty}), but this quantum version has some subtleties. First we have to figure out some missing details in the definition. 

Acts were defined as unitary operators from subspaces of $\HH$ into $\HH$, which can be seen as restrictions of operators acting on the whole $\HH$. As the domain of $V_1$ and $V_2$ was not given, we might take it to be $\HH$. And, as Wallace does not specify a subspace in which the complement of $E$ is to be taken, it seems it should be $\HH$. But then unitarity makes the condition $V_1\rvert_{E^\perp}=V_2\rvert_{E^\perp}$ quite restrictive, forcing $V_1(E)=V_2(E)$. So for the agent to be indifferent between $V_1$ and $V_2$ both must take $E$ to the same range, which does not seem to be what Wallace has in mind. 

More likely, $V_1$ and $V_2$ are intended to be acts available at $\OO_U$, and the complement to be taken with respect to $\OO_U$. So we adopt the following formal definition:

\vspace{6pt}

\begin{Definition}[Null Event]\label{def:Null Event}
Let $M\in\M$, $\psi\in M$, $U\in\U_M$, and $E\in\E$ with $E\subset\OO_U$. Then $E$ is \emph{null} for $\psi$ and $U$ if $V_1U \sim^\psi V_2U$ for any $V_1,V_2\in\U_{\OO_U}$ with $V_1\rvert_{E^\perp\wedge \OO_U}=V_2\rvert_{E^\perp\wedge \OO_U}$.
\end{Definition}

\vspace{6pt}

Wallace claims \cite[p.178]{Wallace2012}, without proof, that finite unions of null sets are null, and subsets of null sets are also null (clearly he means events and disjunctions, instead of sets and unions). So, in formal terms, nullity is to have the following properties, for any $M\in\M$, $\psi\in M$, $U\in\U_M$, and $E, F\in\E$ such that $E, F\subset\OO_U$:
\begin{ThmName}[Null Subevent]\label{pr:Null Subevent}
If $E\subset F$ and $F$ is null for $\psi$ and $U$, then so is $E$.
\end{ThmName}
\begin{ThmName}[Null Disjunction]\label{pr:Null Disjunction}
If $E$ and $F$ are null for $\psi$ and $U$, then so is $E\vee F$.
\end{ThmName}

These have classical counterparts, and the first one is easily proven. The other one seems intuitive: if the agent does not care about his future selves at $E$ or $F$, he should not care about them at $E\vee F$ either. The classical proof uses the fact that classical acts (arbitrary functions mapping states to payoffs) at a classical event (a set of states) and at its complement are independent. It does not adapt to the quantum case, where acts are unitary maps, and what happens at $E$ is linked to what happens at $E^\perp$. An attempt at a direct adaptation might be:
\begin{proof}[Proof (incorrect)]
Let $E$ and $F$ be null for $\psi$ and $U$, and let $V_1, V_2\in \U_{\OO_U}$  satisfy $V_1\rvert_{(E\vee F)^\perp\wedge \OO_U}=V_2\rvert_{(E\vee F)^\perp\wedge \OO_U}$. Without loss of generality, we can assume $E\perp F$, and define a new act $V$ at $\OO_U$ by
\[ V(\phi)=\begin{cases}
V_1(\phi)\ \text{ if } \phi\in E, \\
V_2(\phi)\ \text{ if } \phi\in F, \\
V_1(\phi)=V_2(\phi)\ \text{ if } \phi\in (E\vee F)^\perp \wedge \OO_U.
\end{cases} \]
As $F$ and $E$ are both null, $V_1U \sim^\psi VU \sim^\psi V_2U$.
\end{proof}

The problem with such ``proof'' is that $V$ might not be in $\U_{\OO_U}$, and not even be unitary, as this requires $V_1(E)\perp V_2(F)$, and there is no reason to expect it to hold for all $V_1$ and $V_2$ as above\footnote{This would become even worse if we had defined nullity using $\HH$ instead of $\OO_U$. }.
Wallace would probably argue, using decoherence, that states of $V_1(E)$ and $V_2(F)$ carry records of their distinct pasts, so should be (almost) orthogonal. But none of his axioms formalize such idea. And a new one, stating that under some conditions $E\perp F \Rightarrow V_1(E)\perp V_2(F)$, might conflict with \nameref{ax:Eras}.

\nameref{pr:Null Disjunction} is only used in the \nameref{lm:Nullity Lemma}, which shows $E$ is null for $\psi$ and $U$ if, and only if, $\Pi_E U\psi=0$. So an alternative is to take this as defining nullity:

\vspace{6pt}

\begin{Definition}[Null Event, alternative]
Let $M\in\M$, $\psi\in M$, $U\in\U_M$, and $E\in\E$ such that $E\subset\OO_U$. We say $E$ is \emph{null} for $\psi$ and $U$ if $\Pi_E U\psi=0$.
\end{Definition}

\vspace{6pt}

As $\Pi_E U\psi=0$ means the agent, starting at $\psi$, will have, after $U$, no future selves at $E$, it seems reasonable to adopt such lack of descendants as a characterization of which events he should not care about (i.e. are null). The concept of nullity is only needed for the \nameref{ax:DiacCons} axiom, whose justification in terms of this alternative definition seems as good as the original one.

\subsubsection{Diachronic Consistency}

Wallace's last axiom ``rules out the possibility of a conflict of interest between an agent and his future selves'' \cite[p.168]{Wallace2012}:

\begin{ThmName}[Diachronic Consistency]\label{ax:DiacCons}
Let $M\in\M$, $\psi\in M$, $U\in\U_{M}$ and $V, V'\in\U_{\OO_U}$. Given a partition $\OO_U=\dot{\vee}_i M_i$ with $M_i\in\M$, let  $\phi_i=\Pi_{M_i}U\psi$. Then:
\begin{itemize}
\item if $V\lvert_{M_i}\succcurlyeq^{\phi_i} V'\lvert_{M_i}$ for all $i$ with $M_i$ not null for $\psi$ and $U$, then $VU\succcurlyeq^{\psi} V'U$;
\item if, in addition, $V\lvert_{M_i}\succ^{\phi_i} V'\lvert_{M_i}$ for at least one such $i$, then $VU\succ^{\psi} V'U$.
\end{itemize} 
\end{ThmName}

Here we show that this axiom mixes two ideas, one diachronic and the other synchronic. And that the former follows from the other axioms, so this axiom could be replaced by a simpler one expressing only the second idea.

The diachronic idea corresponds to Wallace's informal description, that preferences should not change in the middle of the decision problem. To isolate it, let $\OO_U$ be a single macrostate $M$. 
In such case the axiom can be reduced to the statement that $V\succ^{U\psi} V' \Leftrightarrow VU \succ^\psi V'U$, which could be obtained from \nameref{ax:StaSup}. 

The synchronic one is, informally, that if $V\lvert_{M_i}$ is preferred, or equivalent, to $V'\lvert_{M_i}$ at all branches, then $V$ is preferred, or equivalent, to $V'$. 
But so far we can not write $V\succ^{U\psi} V'$ when $U\psi$ is branched, as Wallace defined $\succ$ only at macrostates (even though he mentions it would be an order on acts at events \cite[p.166]{Wallace2012}). 

Some new definitions allow us to properly separate the two ideas. 

\vspace{6pt}

\begin{Definition}[Accessible States]
Let $E\in\E$. A $\phi\in E$ is \emph{accessible (from $\psi$, via $U$)} if there are $M\in\M$, $\psi\in M$ and $U\in\U_M$ such that $\OO_U\subset E$ and $U\psi=\phi$.
\end{Definition}

\vspace{6pt}

So a state is accessible if it can be reached from a non-branched state, via some act. Non-branched states are trivially accessible, and, assuming the decision problem starts at a macrostate, all states of interest should be accessible.
We extend $\succ^\psi$, which was only defined at non-branched states, to all accessible ones. 

\vspace{6pt}

\begin{Definition}[Extended Order]\label{def:Extended Order}
Let $\phi\in E$ ($E\in\E$) be accessible from $\psi\in M$ ($M\in\M$) via $U\in\U_M$. For $V,V'\in\U_E$, we write $V\succ^\phi V'$ whenever $VU\succ^\psi V'U$.
\end{Definition}

\vspace{6pt}

By \nameref{ax:StaSup}, $\succ^\phi$ does not depend on $\psi$, $M$ or $U$, and by \nameref{ax:Ord} it is a total order. 
If $E\in\M$ it coincides with the original order. 

We can now state a new axiom, with only the synchronic part of \nameref{ax:DiacCons}. We have incorporated our alternative concept of nullity, but it could be stated in terms of the original one (with small modifications, as we no longer need some $U$ connecting a macrostate $M$ to $E$ just so we can write the preference order).

\begin{ThmName}[Branch Independence]\label{ax:BrIndep}
Let $E\in\E$, $\phi\in E$ be accessible, and $V, V'\in\U_E$. Given a partition $E=\dot{\vee}_i M_i$ with $M_i\in\M$,  let $\phi_i=\Pi_{M_i}\phi$. Then:
\begin{itemize}
\item if $V\lvert_{M_i}\succcurlyeq^{\phi_i} V'\lvert_{M_i}$ for all $i$ with $\phi_i\neq 0$, then $V\succcurlyeq^{\phi} V'$;
\item if, in addition, $V\lvert_{M_i}\succ^{\phi_i} V'\lvert_{M_i}$ for at least one $i$ with $\phi_i\neq 0$, then $V\succ^{\phi} V'$.
\end{itemize} 
\end{ThmName}

It relates the order at a branched state to preferences at its branches, whenever all the agent's versions agree.
If, for all of them, the act corresponding to $V$, via restriction, is preferred, or equivalent, to that given by $V'$, then $V$ is preferred, or equivalent, to $V'$. And the preference is strict if it is so in at least one branch having a version of the agent. This is a many worlds version of the classical \nameref{ax:Indep} axiom (Appendix \ref{sec:Decision under risk}): preference between $V$ and $V'$ is independent of branches in which they coincide, being determined only by branches in which they differ.

So the new idea introduced by Wallace's \nameref{ax:DiacCons} axiom is this independence, not the diachronic one he described and which was already present in \nameref{ax:StaSup}.
But even though our axiom is simpler and could replace his, we will not adopt it, to facilitate comparison with his original presentation.

\subsubsection{Nondegeneracy}\label{sec:Nondegeneracy}

In his proof of the Born Rule Theorem, Wallace dismisses the case of all rewards being equally preferred by saying it renders the theorem trivially true \cite[p.188]{Wallace2012}. That is correct, as it leads all acts to be equally preferred, so by taking $u(r)=0$ for all $r$ we make all expected utilities equal, obtaining the result trivially.

However, in such case the result would be trivial even if the expected utilities had been defined as any other weighted average of the utilities, with no reference to the Born weights. We could even skip the utilities and define $\mathrm{EU}_\psi(U)\equiv 0$ for all $\psi$ and $U$, and the result would still be true for such case.

As Wallace's purpose is not really to represent preferences via expected utilities, but rather to relate them to Born weights, a case solved without them is useless. Of course, one can say this is not the only case, and real life decisions tend to include strict preferences (in another paper, we will argue this is not so simple). If that is so, one should not object to an extra axiom, ensuring at least one strict preference:

\begin{ThmName}[Act Nondegeneracy]\label{ax:ActNDeg}
There are $M\in\M$, $\psi\in M$ and $U,V\in\U_M$ such that $U\succ^\psi V$.
\end{ThmName}

Later, once we obtain the \nameref{def:OrdRe}, we will translate this into a strict preference between at least one pair of rewards (\nameref{lm:ReNonDeg}).

We note that axiom \nameref{ax:S5} of classical decision under uncertainty (Appendix \ref{sec:Decision under uncertainty}) is similar to this one, being necessary because part of the goal there is to obtain subjective probabilities. No such axiom is included in decision under risk (Appendix \ref{sec:Decision under risk}), where the probabilities are given.

\section{Formal Proof}\label{sec:Formal Proof}

The proof was reorganized so some demonstrations could be simplified, but the ideas are essentially the same as in Wallace's. Parallels with classical decision theory are indicated, so its structure can be clarified by comparison with Appendix \ref{sec:Classical Decision Theory}.
Use of axioms is  made explicit, except for \nameref{ax:Restr}, \nameref{ax:Compos},  \nameref{ax:Indol}, \nameref{ax:Cont}, and \nameref{ax:Ord}, which are quite simple and ubiquitous. Many details were filled in to check for problems, and we suggest corrections when these are found.

The first lemma shows preference on bets (represented by acts $U_1, U_2, V_1$ and $V_2$) depends only on the norms of projections of the normalized final states (so, on their Born weights) on reward subspaces. It corresponds, classically, to the assumption, expressed in \eqref{def:lottery}, that in decisions about lotteries all that matters are the rewards and their probabilities. In the Everettian case we are to start with no intuitive meaning for the Born weights, so it takes a few steps to get to the same point.

\begin{ThmName}[Equivalence Lemma]\label{lm:Equivalence Lemma}
Let $M_1,M_2\in\M$, $\psi_1\in M_1$, $\psi_2\in M_2$, $U_1, V_1\in\U_{M_1}$ and $U_2, V_2\in\U_{M_2}$. If, for all $r\in\R$, $\|\Pi_r U_1\psi_1\|/\|\psi_1\|=\|\Pi_r U_2\psi_2\|/\|\psi_2\|$ and $\|\Pi_r V_1\psi_1\|/\|\psi_1\|=\|\Pi_r V_2\psi_2\|/\|\psi_2\|$, then $U_1\succ^{\psi_1} V_1 \Leftrightarrow U_2\succ^{\psi_2} V_2$.
\end{ThmName}

\begin{proof}
Without loss of generality, we can assume $\|\psi_1\|=\|\psi_2\|=1$, with no probabilistic interpretation, just to simplify the notation. 
We divide the proof in 3 steps:

\vspace{6pt}

\textbf{Step 1)} We use branching acts $W_1$ and $ W_2$ to split all branches $\phi_{1,r,i}$ and $\phi_{2,r,j}$ of $U_1\psi_1$ and $U_2\psi_2$, in each  $r$, into new ones $\xi_{1,r,i,j}$ and $\xi_{2,r,j,i}$ of equal norms, still in $r$.

\vspace{6pt}

By the \nameref{ax:OrthCond}, $\Pi_r \OO_{U_1}=\OO_{U_1}\wedge r$ and $\Pi_r \OO_{U_2}=\OO_{U_2}\wedge r$. 
For each $r\in\R$ with $\Pi_r U_1\psi_1\neq 0$, take partitions $\OO_{U_1}\wedge r = \dot{\vee}\!_i\, M_{1,r,i}$ and $\OO_{U_2}\wedge r = \dot{\vee}\!_j\, M_{2,r,j}$ in macrostates, and let $\phi_{1,r,i}=\Pi_{M_{1,r,i}} U_1\psi_1$ and  $\phi_{2,r,j}=\Pi_{M_{2,r,j}} U_2\psi_2$. 

For each $r, i$ and $j$, let $p_{r,i}=\|\phi_{1,r,i}\|^2/\|\Pi_r U_1\psi_1\|^2$ and  $q_{r,j}=\|\phi_{2,r,j}\|^2/\|\Pi_r U_2\psi_2\|^2$. 

By \nameref{ax:BrAv}, there is some $W_1\in\U_{\OO({U_1})}$ such that, for each $r$ and $i$, there is a partition $\OO({W_1\rvert_{M_{1,r,i}}})=\dot{\vee\!}_j\, N_{1,r,i,j}\subset r$ with $\|\xi_{1,r,i,j}\|^2=\|\phi_{1,r,i}\|^2\cdot q_{r,j}$, where
$\xi_{1,r,i,j}=\Pi_{N_{1,r,i,j}} W_1\phi_{1,r,i}$.
Likewise, we get $W_2\in\U_{\OO({U_2})}$ such that, for each $r$ and $j$, there is a partition $\OO({W_2\rvert_{M_{2,r,j}}})=\dot{\vee\!}_i\, N_{2,r,j,i}\subset r$ with ${\|\xi_{2,r,j,i}\|^2=p_{r,i}\cdot\|\phi_{2,r,j}\|^2}$, where 
$\xi_{2,r,j,i}=\Pi_{N_{2,r,j,i}} W_2\phi_{2,r,j}$.

As $\|\Pi_r U_1\psi_1\|=\|\Pi_r U_2\psi_2\|$, we have 
\begin{equation*} 
\|\xi_{1,r,i,j}\| 
		 = \frac{\|\phi_{1,r,i}\| \cdot \|\phi_{2,r,j}\|}{\|\Pi_r U_2\psi_2\|} 
		 = \|\xi_{2,r,j,i}\|.
\end{equation*}

\vspace{6pt}

\textbf{Step 2)} We ensure the $N_{1,r,i,j}$'s (resp. $N_{2,r,j,i}$'s) are mutually orthogonal, and use erasures $X_1$ and $X_2$ to make $U_1\psi_1$ and $U_2\psi_2$ end up in the same final state.

\vspace{6pt}

By \nameref{ax:Irrev}, the $\OO({W_1\rvert_{M_{1,r,i}}})$'s are mutually orthogonal, as the $\OO({W_2\rvert_{M_{2,r,j}}})$'s. Hence
$\OO_{W_1} = {\dot{\vee}}\!_{r,i,j}\, N_{1,r,i,j}$ and $\OO_{W_2} = {\dot{\vee}}\!_{r,j,i}\, N_{2,r,j,i}$.
Also, $\xi_{1,r,i,j}=\Pi_{N_{1,r,i,j}}W_1U_1\psi_1$,
and $\xi_{2,r,j,i}=\Pi_{N_{2,r,j,i}}W_2U_2\psi_2$. 

\nameref{ax:Eras} gives $X_1\in\U_{\OO({W_1})}$ and $X_2\in\U_{\OO({W_2})}$ with $X_1(N_{1,r,i,j}), X_2(N_{2,r,j,i})\subset r$ and $X_1 \xi_{1,r,i,j} = X_2 \xi_{2,r,j,i}$.
Then $X_1W_1U_1\psi_1=X_2W_2U_2\psi_2$.

\vspace{6pt}

\textbf{Step 3)} The agent is indifferent to the $W$'s and $X$'s, which take the $U\psi$'s to the same final state. As a similar procedure can be done to the $V\psi$'s, and the agent only cares about the final states, preferences between the $U$'s and the $V$'s must agree.

\vspace{6pt}

By \nameref{ax:BrIndif} we have 
$W_1\rvert_{M_{1,r,i}} \sim^{\phi_{1,r,i}} \mathds{1}_{M_{1,r,i}}$,
$W_2\rvert_{M_{2,r,j}} \sim^{\phi_{2,r,i}} \mathds{1}_{M_{2,r,j}}$, 
$X_1\rvert_{N_{1,r,i,j}} \sim^{\xi_{1,r,i,j}} \mathds{1}_{N_{1,r,i,j}}$ and 
$X_2\rvert_{N_{2,r,j,i}} \sim^{\xi_{2,r,j,i}} \mathds{1}_{N_{2,r,j,i}}$, so \nameref{ax:DiacCons} gives  $X_1W_1U_1\sim^{\psi_1} U_1$ and $X_2W_2U_2\sim^{\psi_2} U_2$.

Writing $U'_1=X_1 W_1$ and $U'_2=X_2 W_2$, we have $U'_1 U_1 \sim^{\psi_1} U_1$ and $U'_2 U_2 \sim^{\psi_2} U_2$, and, from the 2\textsuperscript{nd} step, $U'_1 U_1 \psi_1 = U'_2 U_2 \psi_2$.
Likewise, we get  $V'_1\in\U_{\OO({V_1})}$ and $V'_2\in\U_{\OO({V_2})}$  with  $V'_1 V_1 \sim^{\psi_1} V_1$, \ $V'_2 V_2 \sim^{\psi_2} V_2$ and $V'_1 V_1\psi_1=V'_2 V_2\psi_2$.
By \nameref{ax:StaSup} $U'_1 U_1\succ^{\psi_1}V'_1 V_1 \Leftrightarrow U'_2 U_2\succ^{\psi_2}V'_2 V_2$, so \nameref{ax:Ord} gives $U_1\succ^{\psi_1}V_1 \Leftrightarrow U_2\succ^{\psi_2} V_2$.
\end{proof}

\begin{remark}
There is a small problem in this proof, as some $p_{r,i}$ or $q_{r,j}$ could be $0$ (in Wallace's version \cite[p.183]{Wallace2012}, it corresponds to numbers in his $\mathcal{P}_r$ sets being $0$).
As $M_{1,r,i}\subset \OO_{U_1}$, there is some $\psi'\in M_1$ such that $\Pi_{M_{1,r,i}} U_1\psi'\neq 0$, so intuitively we might expect that $\Pi_{M_{1,r,i}} U_1\psi_1\neq 0$, since $\psi_1$ and $\psi'$ are in the same macrostate. But such idea is not formalized by Wallace's axioms. A way to deal with this is to allow for nonnegative instead of positive numbers in \nameref{ax:BrAv} (this might require some care in its physical justification), and to remove the requirement that states be nonzero in \nameref{ax:Eras} (the axiom is actually trivial if $\psi_i =\phi_i=0$).
\end{remark}
\begin{remark}
Wallace includes \nameref{ax:ReAv} and \nameref{ax:MacIndif} among the hypotheses of this lemma \cite[p.182]{Wallace2012}, but they do not seem to be used anywhere in his demonstration (nor in ours).
\end{remark}

\begin{Corollary}\label{cor:EquivalenceSim}
Let $M\in\M$, $\psi\in M$, and $U,V\in\U_M$. If $\|\Pi_r U\psi\|=\|\Pi_r V\psi\|$ for all $r\in\R$ then $U\sim^\psi V$.
\end{Corollary}

\vspace{6pt}

\begin{Definition}[Single Reward Acts]
Given $r\in\R$, a \emph{single reward act $V_r$ at $E\in\E$} is any $V_r\in\U_E$ such that  $\OO_{V_r}\subset r$.
\end{Definition}

\vspace{6pt}

\nameref{ax:ReAv} ensures such acts are always available. They give the same reward in all branches, corresponding to the classical single reward lotteries.
Classically, an order on rewards is obtained by identifying such lotteries and their rewards.
The \nameref{lm:Equivalence Lemma} allows us to do the same here, ensuring the following definition does not depend on the choice of $V_r, V_s,M$ or $\psi$.

\vspace{6pt}

\begin{Definition}[Order on Rewards]\label{def:OrdRe}
Given $r, s\in\R$, we write $r\succ s$ whenever $V_r\succ^\psi V_s$ for single reward acts $V_r$ and $V_s$ at some $M\in\M$, and some $\psi\in M$. 
\end{Definition}

\vspace{6pt}

From $\succ$, the symbols $\prec$, $\sim$, $\succcurlyeq$,  and $\preccurlyeq$ are defined as usual. 
By \nameref{ax:ReAv} and \nameref{ax:Ord} this is a total order on $\R$.
As noted in section \ref{sec:Macrostate Indifference}, Wallace claims to obtain it before the lemma, which he uses instead to get an order on reward functions. 

As seen in section \ref{sec:Nondegeneracy}, he dismisses the case of $r\sim s$ for all $r,s\in\R$ by saying it renders his result trivial. We instead obtain the nondegeneracy of rewards from that on acts, since the order on rewards is derived from the one on acts. 

\begin{ThmName}[Reward Nondegeneracy Lemma]\label{lm:ReNonDeg}
There are $r, s\in\R$ such that $r \succ s$. 
\end{ThmName}
\begin{proof}
Fix $s\in\R$, and suppose $r \sim s$ for all $r\in\R$. \nameref{ax:ActNDeg} gives $M\in\M$, $\psi\in M$ and $U,V\in\U_M$ such that $U\succ^\psi V$. 
After partitioning each $\OO_U\wedge r$ ($r\in\R$) into macrostates, we can use \nameref{ax:ReAv} to get $U'\in\U_{\OO_U}$ such that $\OO_{U'}\subset s$. 
The \nameref{def:OrdRe} and
\nameref{ax:DiacCons} imply $U'U\sim^\psi U$. 
Likewise, we get $V'\in\U_{\OO_V}$ with $\OO_{V'}\subset s$ and $V'V\sim^\psi V$.
By \autoref{cor:EquivalenceSim}, $U'U\sim^\psi V'V$, so that $U\sim^\psi V$, contradicting their choice. 
\end{proof}

\begin{Definition}[Extremal Rewards]
We fix $r_0, r_1\in\R$ such that $r_0\prec r_1$ and $r_0\preccurlyeq r \preccurlyeq r_1$ for all $r\in\R$.
\end{Definition}

\vspace{6pt}

Existence of $r_0$ and $r_1$ is assured by the previous lemma and the finiteness of $\R$. As stated in section \ref{sec:Rewards}, we assume $\R$ is finite just for simplicity, and the proof can be adapted to work without extremal rewards. But it would require some technical workarounds which, though usual in the classical theory and present in Wallace's original proof, would cloud the main ideas without bringing any relevant gain.

The next lemma characterizes a \nameref{def:Null Event}, showing the only way an agent will not care about an event is if none of his versions are in it. This is the only part of the proof whose classical parallel is in decision under uncertainty (Appendix \ref{sec:Decision under uncertainty}). It comes into play here because, instead of lotteries, Wallace uses a framework of acts and events, typical of that case.

\begin{ThmName}[Nullity Lemma]\label{lm:Nullity Lemma}
Let $M\in\M$, $\psi\in M$, $U\in\U_M$, and $E\in\E$ such that $E\subset\OO_U$. Then $E$ is null for $\psi$ and $U$ if, and only if, $\Pi_E U\psi=0$.
\end{ThmName}

\begin{proof}
If $\Pi_E U\psi=0$ then $V_1\rvert_{E^\perp\wedge \OO_U}=V_2\rvert_{E^\perp\wedge \OO_U}$ implies $V_1U\psi=V_2U\psi$, and by \nameref{ax:StaSup} $V_1U\sim^\psi V_2U$. Hence $E$ is null.
The converse consists of 2 steps:

\vspace{6pt}

\textbf{Step 1)} We show the nullity of an event $E$ for a state $\psi$ and an act $U$ is solely determined by the value of $\|\Pi_E U\psi\|/\|\psi\|$.

\vspace{6pt}

Given any $M$, $\psi$, $U$, $E$ as in the axiom, with \nameref{ax:ReAv} we can obtain $V_0, V_1\in\U_{\OO_U}$ such that $\OO_{V_0}\subset r_0$, $V_1\rvert_{E^\perp\wedge \OO_U}=V_0\rvert_{E^\perp\wedge \OO_U}$, and $V_1(E)\subset r_1$. 
If $E$ is null for $\psi$  and $U$  then $V_1U\sim^\psi V_0U$, otherwise,  the \nameref{def:OrdRe}  and \nameref{ax:DiacCons} imply $V_1U\succ^{\psi}V_0U$.

As the same holds for any other $M'$, $\psi'$, $U'$, $E'$ as above,  the \nameref{lm:Equivalence Lemma} implies that, if $\|\Pi_E U\psi\|/\|\psi\|=\|\Pi_{E'}U'\psi'\|/\|\psi'\|$, then $E$ and $E'$ are either both null or both non null.

\vspace{6pt}

\textbf{Step 2)} We show that a null $E$ with $\Pi_E U\psi\neq 0$ leads to a contradiction.

\vspace{6pt}

Suppose there are $M$, $\psi$, $U$, $E$ as in the axiom, with $E$ null for $\psi$ and $U$ but ${\|\Pi_E U\psi\|^2/\|\psi\|^2=w>0}$.
Taking any $n\in\mathbb{N}$ with $1/n<w$, 
\nameref{ax:BrAv} gives $V\in\U_{M}$ and a partition $\OO_{V}=M_1\dot{\vee} M_2\dot{\vee} M_3$ in macrostates with
\[ \|\Pi_{M_1}V\psi\|^2/\|\psi\|^2= 1/n, \ \ \|\Pi_{M_2}V\psi\|^2/\|\psi\|^2 = w-1/n, \ \ \|\Pi_{M_3}V\psi\|^2/\|\psi\|^2 = 1-w.\]
As $\|\Pi_{M_1\dot{\vee} M_2}V\psi\|^2 = w$, $M_1\dot{\vee} M_2$ is null for $\psi$ and $V$. By \nameref{pr:Null Subevent}, so is $M_1$. 
Hence whenever $\|\Pi_{E}U\psi\|^2/\|\psi\|^2  = 1/n$, $E$ will be null for $\psi$ and $U$.

\nameref{ax:BrAv} gives $W\in\U_M$ and a partition $\OO_{W}=M_1\dot{\vee} \ldots\dot{\vee} M_n$ with $\|\Pi_{M_i}W\psi\|^2 = 1/n$ for all $i$. As the $M_i$'s are null for $\psi$ and $W$,  by \nameref{pr:Null Disjunction} so is $\OO_{W}$. 
So whenever $U\psi\in E$, $E$ will be null for $\psi$ and $U$.

Given $M\in\M$ and $\psi\in M$, $M$ will be null for $\psi$ and $\mathds{1}_M$. 
This contradicts the \nameref{def:OrdRe}, which gives $V_{r_1}\succ^{\psi}V_{r_0}$ for single reward acts $V_{r_0}$ and $V_{r_1}$ at $M$.
\end{proof}

\begin{remark}
This lemma might be invalidated if the problem with \nameref{pr:Null Disjunction} is not solved. But this might not jeopardize the rest of the proof, if \nameref{ax:DiacCons} is accepted with the alternative definition of nullity discussed in section \ref{sec:Nullity}.
\end{remark}

We now define a quantum version of the standard lotteries \eqref{eq:standard lottery}.
They play in our presentation a role similar to the reward functions $f[\alpha]$ Wallace defines in his Dominance Lemma \cite[p.185]{Wallace2012} (in his terminology, with $t=r_0$ and $s=r_1$, our standard act $U_w$ corresponds to an act whose reward function is $f[w]$).

\vspace{6pt}

\begin{Definition}[Standard Acts]\label{def:standard act}
Let $M\in\M$, $\psi\in M$ and $0\leq w\leq1$. A \emph{standard act of weight $w$ at $\psi$} is any $U_w\in\U_M$ with $\|\Pi_{r_1} U_w\psi\|^2/\|\psi\|^2=w$ and $\|\Pi_{r_0} U_w\psi\|^2/\|\psi\|^2=1-w$. 
\end{Definition}

\vspace{6pt}

By \nameref{ax:BrAv} and \nameref{ax:ReAv}, these are always available.

The next result corresponds to step \ref{step:monotonicity} in the proof of the \nameref{th:VNM}  (Appendix \ref{sec:Decision under risk}). 
In Wallace's original presentation, it is stated in terms of reward functions $f[\alpha]$ instead of standard acts, but the idea is the same.

\begin{ThmName}[Dominance Lemma]\label{lm:Dominance Lemma}
Let $M\in\M$, $\psi\in M$ and $w,w'\in[0,1]$. If $U_{w}$ and $U_{w'}$ are standard acts of weights $w$ and $w'$ at $\psi$ then $w'>w \Leftrightarrow U_{w'}\succ^\psi U_{w}$.
\end{ThmName}
\begin{proof}
If $w'=w$ then $U_{w'}\sim^\psi U_{w}$, by \autoref{cor:EquivalenceSim}. 

If $w'>w$ then, using \nameref{ax:BrAv}, we can obtain $U\in\U_M$ and a partition $\OO_U=M_1\dot{\vee} M_2 \dot{\vee} M_3$ in macrostates such that
\[ \|\Pi_{M_1}U\psi\|^2/\|\psi\|^2 = w, \quad
\|\Pi_{M_2}U\psi\|^2/\|\psi\|^2 = w'-w, \quad
\|\Pi_{M_3}U\psi\|^2/\|\psi\|^2 = 1-w'. \] 
\nameref{ax:ReAv} gives $V,V'\in\U_{\OO_U}$ with 
\begin{align*}
V(M_1)&\subset r_1, & V(M_2)&\subset r_0, & V(M_3) &\subset r_0, \\
V'(M_1)&\subset r_1, & V'(M_2)&\subset r_1, & V'(M_3) &\subset r_0,
\end{align*}
so that, by the \nameref{def:OrdRe}, 
\[V'\rvert_{M_1} \sim^{\Pi_{M_1}U\psi} V\rvert_{M_1}, \qquad V'\rvert_{M_2} \succ^{\Pi_{M_2}U\psi} V\rvert_{M_2}, \qquad V'\rvert_{M_3} \sim^{\Pi_{M_3}U\psi} V\rvert_{M_3}.\] 
By the \nameref{lm:Nullity Lemma}\footnote{Or our alternative definition of nullity.}, $M_2$ is not null for $\psi$ and $U$, so \nameref{ax:DiacCons} gives $V'U\succ^\psi VU$. 
By \autoref{cor:EquivalenceSim}, $V'U\sim^\psi U_{w'}$ and $VU\sim^\psi U_w$, so that $U_{w'}\succ^\psi U_{w}$.

The converse follows via \nameref{ax:Ord}.
\end{proof}

The rest of the proof, though based on Wallace's ideas, is organized in quite a different way than his, so some arguments could be simplified, and to help clarify its structure. 

The next lemma, corresponding to step \ref{step:continuity} in the proof of the \nameref{th:VNM}, forms an \emph{utility function} $u(r)$ by comparing single reward acts $V_r$ with standard acts $U_w$. It is similar to Wallace's Utility Lemma, but differs in the extent to which it develops properties of $u$ (hence the different name). Part of Wallace's lemma was separated into our remaining results, while part of the demonstration of his Born Rule Theorem was incorporated into this lemma.

\begin{ThmName}[Utility Function Lemma]\label{lm:Utility Function Lemma}
There is an unique\footnote{Other values of $u(r_0)$ or $u(r_1)$ allow for positive affine transformations, as in the classical theory.}  $u:\R\rightarrow [0,1]$ such that:
\begin{itemize}
\item $u(r_0)=0$ and $u(r_1)=1$;
\item for any $r\in\R$, $M\in\M$ and $\psi\in M$, we have $V_r\sim^\psi U_{u(r)}$, where $V_r$ is a single reward act at $M$, and $U_{u(r)}$ is a standard act of weight $u(r)$ at $\psi$.
\end{itemize}
Moreover, $u(r)>u(s) \Leftrightarrow r\succ s$.
\end{ThmName}
\begin{proof}
For any $r\in\R$, $M\in\M$, $\psi\in M$, and $0\leq w\leq 1$, let $V_r$ be a single reward act at $M$, and $U_w$ a standard act of weight $w$ at $\psi$. 
By the \nameref{lm:Equivalence Lemma}, the definition
\begin{equation*}
u(r)=\sup\{w:U_w \preccurlyeq^\psi V_r\},
\end{equation*}
does not depend on the choice of $\psi, M, V_r$ or $U_w$.
By construction, we have $U_w \succ^\psi V_r$ if $w > u(r)$, and by the \nameref{lm:Dominance Lemma} $U_w \prec^\psi V_r$ if $w<u(r)$.

Suppose $U_{u(r)} \succ^\psi V_r$.
Then $u(r)\neq 0$, or the \nameref{def:OrdRe} would give $r_0\succ r$. 
By \nameref{ax:SolCont}, $U\succ^\psi V_r$ for all $U$ in a neighborhood of $U_{u(r)}$. 
For $w<u(r)$ close enough to $u(r)$, we can assume 
$U_{w}$ is in this neighborhood (if necessary, using \nameref{ax:PrCont} to get $U_{w}$ as a perturbation of $U_{u(r)}$), contradicting $U_w \preccurlyeq^\psi V_r$. 
By a similar argument it is not possible that $U_{u(r)} \prec^\psi V_r$.
So $U_{u(r)} \sim^\psi V_r$, and by the \nameref{lm:Equivalence Lemma} this holds for any other choices of $V_r, U_{u(r)}$ and $\psi$. 

The \nameref{lm:Dominance Lemma} and the \nameref{def:OrdRe} imply that 
\[ u(r)>u(s) \ \Leftrightarrow\   U_{u(r)} \succ^\psi U_{u(s)} 
\ \Leftrightarrow\ V_r \succ^\psi  V_s \ \Leftrightarrow\ r\succ s. \]

To prove unicity, let $\tilde{u}$ be any other function with the same properties. Then $U_{\tilde{u}(r)}\sim^\psi V_r\sim^\psi U_{u(r)}$, and by the \nameref{lm:Dominance Lemma} $\tilde{u}(r)=u(r)$.
\end{proof}

\begin{remark}
In his proof of the Born Rule Theorem \cite[p.188]{Wallace2012}, Wallace uses \nameref{ax:BrAv} and \nameref{ax:ReAv} to get an act $U_{i,\alpha}$ in a neighborhood $\mathcal{N}_i$ of another act $U_i$, with slightly different reward functions. Those axioms do give an act with the desired reward function, but it might be nowhere close to $U_i$. To ensure it is in $\mathcal{N}_i$ one must instead use \nameref{ax:PrCont}, as we did in the proof above.
\end{remark}

\begin{Definition}[Born Average Utility]
The \emph{Born average utility} of an act $U$ at $\psi$ is 
\begin{equation*}
\mathrm{BU}_\psi(U) = \sum_{r\in\R} \dfrac{\|\Pi_r U\psi\|^2}{\|\psi\|^2}\cdot u(r).
\end{equation*}
\end{Definition}

This is an average of the utilities of the rewards, weighted by the Born weights of $U\psi$ on them.
In Wallace's terminology, it is the expected utility $\mathrm{EU}_\psi(U)$, corresponding classically to \eqref{eq:Expected Utility Classical}, but with Born weights in place of probabilities, as in Deutsch's proposal \eqref{eq:Expected Utility in Deutsch}. 
As the term \emph{expected} evoques its usual meaning as a probability-weighted average, and at this point any link between Born weights and probabilities is yet to be established, we adopt a more neutral terminology.

For standard acts, $\mathrm{BU}(U_w)=w$.
The next lemma shows any act is equivalent to a standard act with the same Born average utility. It corresponds to step \ref{step:substitutability} in the proof of the \nameref{th:VNM}. 

\begin{ThmName}[Standard Act Lemma]\label{lm:Standard Act Lemma}
Let $M\in\M$, $\psi\in M$, $U\in\U_M$. Then $U\sim^\psi U_w$, where $U_w$ is a standard act of weight $w=\mathrm{BU}_\psi(U)$ at $\psi$.
\end{ThmName}

\begin{proof}
For each $r\in\R$ let $\{M_{r,i}\}$ be a partition of $\Pi_r\OO_U$ in macrostates, and let $\phi_{r,i}=\Pi_{M_{r,i}}U\psi$. 
With \nameref{ax:BrAv} and \nameref{ax:ReAv} we can get $W\in\U_{\OO_U}$ such that $W\lvert_{M_{r,i}}$ is a standard act of weight $u(r)$ at $\phi_{r,i}$ whenever $\phi_{r,i}\neq 0$. 
Then $\|\Pi_{r_1}W\phi_{r,i}\|^2=u(r)\cdot\|\phi_{r,i}\|^2$, and as by \nameref{ax:Irrev} the $\Pi_{r_1}W\phi_{r,i}$'s are mutually orthogonal, we have
\begin{equation*} 
\|\Pi_{r_1}WU\psi\|^2 
	=\sum_{r,i}\left\|\Pi_{r_1}W\phi_{r,i}\right\|^2  
	= \sum_r u(r)\cdot\|\Pi_r U\psi\|^2 = 
	\mathrm{BU}_\psi(U) \cdot \|\psi\|^2.
\end{equation*}
Hence $WU$ is a standard act of weight $\mathrm{BU}_\psi(U)$ at $\psi$.
The \nameref{lm:Utility Function Lemma} implies $W\lvert_{M_{r,i}}\sim^{\phi_{r,i}} \mathds{1}_{M_{r,i}}$, so \nameref{ax:DiacCons} gives $WU\sim^\psi U$.
\end{proof}

Finally, as preference on standard acts increases with their Born average utilities, the same holds for all other acts. This is corresponds to the last step in the proof of the \nameref{th:VNM}, leading to its quantum equivalent.

\begin{ThmName}[Wallace Theorem]\label{th:Wallace Theorem}
If $\succ^\psi$ is a Wallacean solution to a rich quantum decision problem, then there is an utility function $u$ on rewards that represents it via a \emph{Principle of Maximization of Born Average Utility}, i.e.
\[U\succ^\psi V \quad\Leftrightarrow\quad \mathrm{BU}_\psi(U)>\mathrm{BU}_\psi(V).\]
\end{ThmName}
\begin{proof}
By the \nameref{lm:Standard Act Lemma}, $U$ and $V$ are equivalent to standard acts of weights $\mathrm{BU}_\psi(U)$ and $\mathrm{BU}_\psi(V)$ at $\psi$, so the \nameref{lm:Dominance Lemma} gives the result.
\end{proof}

Wallace calls this the Born Rule Theorem. We avoid such name as its connection with the Born Rule is not clear at this point. 
Granted, there is a parallel with the classical \nameref{th:PMEU}, except for the use of Born weights instead of probabilities in the weighted average of utilities:
\begin{itemize}
\item Everettian agents decide using Born weights as parameters to measure the relevance of sets of branches;
\item Classical agents decide using probabilities as parameters to measure the relevance of possible alternatives. 
\end{itemize}  

But to say that the first parameters acquire some meaning by comparison with the second case would be precipitate. Not all relevance is probabilistic, if a teacher adopts a weighted grade it does not mean he thinks one test is more likely to happen than another. H.\,Greaves \cite{Greaves2004,Greaves2007} has even suggested interpreting Born weights not as probabilities, but as a \emph{caring measure} that quantifies how much the agent should care about each of his future selves.

And, as tempting as it may be to consider any illusion of chance in EQM in terms of Savage's subjective probabilities, we note that Born weights have objective values imposed by the theory. 
And as our presentation shows, Wallace's proof seems closer to classical decision under risk than to Savage's work.

So there is still a long way to go from this result to a probabilistic interpretation of the Born weights. In \cite{Wallace2012} Wallace goes to great lengths in an attempt to establish such link. As such interpretation falls out of the scope of this article, this discussion will be left for another one.

\section{Conclusion}\label{sec:Conclusion}

From a purelly formal perspective, Wallace's proof seems to be, for the most part, correct. Still, a few problems were found:
\begin{itemize}
\item \nameref{pr:Null Disjunction} seems incorrect, invalidating the \nameref{lm:Nullity Lemma}. But the lemma can be skipped if \nameref{ax:DiacCons} is valid with our definition of nullity.

\item The way Wallace suggests the \nameref{def:OrdRe} can be obtained does not seem to work, requiring instead the \nameref{lm:Equivalence Lemma}. 

\item The statements of \nameref{ax:Irrev} and \nameref{ax:Eras} lacked important hypotheses.

\item Some corrections were needed in the concepts of event and $\mathcal{P}$-branching, and in the axioms of \nameref{ax:BrAv} and \nameref{ax:Eras}.

\item A small problem in the proof of the \nameref{lm:Equivalence Lemma} might require other changes in \nameref{ax:BrAv} e \nameref{ax:Eras}.

\item \nameref{ax:PrCont} must be used to get $U_{i,\alpha}$ in Wallace's demonstration of his Born Rule Theorem, instead of \nameref{ax:BrAv} and \nameref{ax:ReAv}.
\end{itemize}
If all problems can be corrected, which seems plausible, the proof should be formally valid. We have also identified some possible improvements:
\begin{itemize}
\item If all rewards are equivalent, Wallace's result becomes so trivial that Born weights lose their relevance. A new axiom was added to avoid this.

\item \nameref{ax:Indol} and \nameref{ax:Cont} could be replaced by one simpler axiom.

\item The statement of \nameref{ax:MacIndif} does not seem to correspond to what Wallace intended the axiom to be. Anyway, it is not needed for the proof.

\item \nameref{ax:DiacCons} could be replaced by a simpler axiom, as \nameref{ax:StaSup} implies part of it.

\item The proof is simpler for finitely many rewards. For Wallace's purpose, this should not be a relevant loss of generality.

\item Changing the order results are proven allowed some arguments to be simplified.
\end{itemize}

Our analysis also shows that the main ingredient for the proof is provided by the \nameref{lm:Equivalence Lemma}. It shows that, for quantum decisions, all that matters are the Born weights in reward subspaces, just as classical lotteries are characterized by the probabilities of rewards. So those weights, which did not seem to play any role in Everettian Quantum Mechanics, end up being determinant factors in quantum decision problems. Having this, the rest of the proof is just a quantum version of standard arguments of decision theory.

Some questions still demand careful examination before Wallace's proof can be seen as solving the probability problem. For one, the result of a formal proof is only as good as the axioms and concepts upon which it is based. Are they physically reasonable? Can they really be seen as characterizing rationality for Everettian agents?

Another question is how to interpret the result. Are Born weights an Everettian agent's subjective probabilities? They have objective values, the agent is supposed to know them, evolution is deterministic, and all branches exist. So there does not seem to be much room left for subjectivity or uncertainty.
Despite this, can some probabilistic meaning still be attributed to the Born weights?  

Wallace and other authors have dedicated much thought to these questions, but analyzing their arguments is a discussion beyond the aims of this article.
Only after such doubts are settled can the use of terms like rationality axioms, rational strategy, Born Rule Theorem or expected utility be justified. Our preference for a more neutral terminology, such as Wallacean solution, Wallace Theorem and Born average utility, stems from the fact that much of the debate about these questions has been clouded by the use of value-laden terms.

We hope our analysis of Wallace's proof helps shed light on the principles and arguments involved, and on what its result really means. This article is actually intended as laying the groundwork for other papers dealing with those questions.

\appendix

\section{Classical decision theory}\label{sec:Classical Decision Theory}

Decision Theory \cite{Karni2014,Kreps1988,Parmigiani.2009} is an interdisciplinary area of study, concerned, among other things, with developing principles to optimize decision making.
Its problems are usually classified in 3 cases (the nomenclature varies in the literature): decision under certainty, under risk, and under uncertainty.

\subsection{Decision under certainty}\label{sec:Decision under certainty}

This case involves choosing between alternatives with well known outcomes. Such problems can be framed in terms of an agent, or decision maker, having to establish a preference order on acts whose outcomes are certain (sometimes they are framed in terms of lotteries with a single reward).

This preference is called \emph{rational} if it is a total order, i.e. it satisfies \nameref{ax:Compl} and \nameref{ax:Trans} (see next section for definitions), so the choices can be ordered in a single chain without loops.
The reason for such label is that an agent whose preference order is not total would be vulnerable to Dutch book arguments (loops in the order could be exploited to ``pump money'' from him).

\subsection{Decision under risk}\label{sec:Decision under risk}

In this kind of problem, each choice can have multiple outcomes, with known probabilities. Such problems are usually framed in terms of an agent having to establish a \emph{preference order} $\succ$ on lotteries. A \emph{lottery} 
\begin{equation}\label{def:lottery}
A=\{(p_i, r_i)\}_{i\in\mathcal{O}}
\end{equation}
consists of a set $\mathcal{O}$ of mutually exclusive \emph{outcomes}, each having probability $p_i$ and giving a \emph{reward} $r_i$ (which can also be a penalty or any other consequence). 

Each reward $r$ is identified with a \emph{single reward lottery}, giving it with certainty as its unique reward. So an order on lotteries induces another on rewards. 

Given lotteries $A$ and $B$, and $t\in[0,1]$, a \emph{compound lottery} $tA+(1-t)B$ is a lottery in which the agent first chooses randomly $A$ or $B$, with probabilities $t$ and $1-t$ respectively, and then proceeds with the selected lottery. By classical probability rules, it is equivalent to one in which the probability of each reward is $tp_A+(1-t)p_B$, where $p_A$ and $p_B$ are its probabilities in $A$ and $B$. 

If all rewards are equally preferred, the problem is trivial. 
Otherwise, we assume\footnote{This is just for simplicity, the theory can be adapted to work without such extremal rewards.} the existence of least and most preferred rewards, denoted, respectively, by $r_0$ and $r_1$. For each $t\in[0,1]$, we define a \emph{standard lottery of weight $t$} as
\begin{equation}\label{eq:standard lottery}
L_t=tr_1+(1-t)r_0.
\end{equation}

A crude strategy for problems of decision under risk is to let preferences follow the expected value of lotteries, i.e. the average value of rewards, weighted by their probabilities.
This may seem reasonable, by the Law of Large Numbers, but can lead to bad results, as in the St. Petersburg Paradox, if the number of runs is finite. 
A more flexible strategy, proposed in 1738 by D.\,Bernoulli \cite{Bernoulli.1954}, replaces monetary values by an \emph{utility function} $u(r)$ on rewards:

\begin{ThmName}[Principle of Maximization of Expected Utility]\label{th:PMEU} 
Given an utility function $u(r)$, it induces a preference order on lotteries by
$A \succ B \Leftrightarrow \mathrm{EU}(A) > \mathrm{EU}(B)$,
where the \emph{expected utility} of  $A=\{(p_i,r_i)\}_{i\in\mathcal{O}}$ is
\begin{equation}\label{eq:Expected Utility Classical}
\mathrm{EU}(A) = \sum_{i\in\mathcal{O}} p_i \cdot u(r_i).
\end{equation}
\end{ThmName}

An order $\succ$ on lotteries is induced (or \emph{represented}) by a function $u$ on rewards, via this principle, if, and only if, the following conditions are satisfied, for any rewards $r$, $r'$ and $r''$:
\begin{itemize}
\item  $r\succ r' \ \Leftrightarrow \ u(r)>u(r')$;
\item if $r'' \sim tr+(1-t)r'$ then $u(r'')=t\cdot u(r)+(1-t)\cdot u(r')$.
\end{itemize}

The importance of this principle became clear in 1944, when Von Neumann and Morgenstern \cite{Neumann1944} proved that any preference order satisfying four reasonable conditions can be represented by an utility function. The \emph{Von Neumann-Morgenstern axioms} are, for any lotteries $A, B, C$:
\begin{ThmName}[Completeness]\label{ax:Compl} 
Either $A\succ B$, or $B\succ A$, or $A\sim B$.
\end{ThmName}
\begin{ThmName}[Transitivity]\label{ax:Trans}
If $A\succcurlyeq B$ and $B\succcurlyeq C$ then $A\succcurlyeq C$.
\end{ThmName}
\begin{ThmName}[Archimedean Property]\label{ax:Arch}
If $A\succ B\succ C$, there are $s,t\in(0,1)$ such that
$tA+(1-t)C \succ B \succ sA+(1-s)C$.
\end{ThmName} 
\begin{ThmName}[Independence]\label{ax:Indep}
If $A\succ B$ then  $tA+(1-t)C \succ tB+(1-t)C$, for any $t\in(0,1]$.
\end{ThmName}
\nameref{ax:Compl} and \nameref{ax:Trans} correspond to the totality condition of decision under certainty.
The \nameref{ax:Arch} implies no lottery is so incommensurately better (resp. worse) than other, that it is impossible to reverse preferences by compounding it appropriately with a worse (resp. better) one. It also means that sufficiently small changes in a lottery do not alter significantly the order. 
\nameref{ax:Indep} means a preference between compound lotteries is based only on components that differ. 

The first three axioms ensure $\succ$ can be described by a correspondence between lotteries and real numbers. With the last one, they also imply the following properties:
\begin{ThmName}[Substitutability]\label{ax:Subst}
If $A\sim B$ then $tA+(1-t)C \sim tB+(1-t)C$, for any $t\in[0,1]$; 
\end{ThmName}
\begin{ThmName}[Monotonicity]\label{ax:Monot}
If $A\succ B$ then $tA+(1-t)B$ is more preferable for higher values of $t$;
\end{ThmName}
\begin{ThmName}[Continuity]\label{ax:Continuity}
If $A\succ B\succ C$ there is a unique $t\in(0,1)$ such that $B\sim tA+(1-t)C$.
\end{ThmName}

From them, we can get the following result, whose proof we sketch for comparison with Wallace's.

\begin{ThmName}[Von Neumann-Morgenstern Theorem]\label{th:VNM}
A preference order $\succ$ satisfies the Von Neumann-Morgenstern axioms if, and only if, there is an utility function $u$\footnote{Unique up to positive affine transformations $u\mapsto au+b$, for real constants $a>0$ and $b$.} that represents it via the Principle of Maximization of Expected Utility, i.e.
\[A \succ B \quad\Leftrightarrow\quad \mathrm{EU}(A) > \mathrm{EU}(B).\]
\end{ThmName}
\begin{proof}
It is easy to show that any $\succ$ represented by some $u$ satisfies the axioms. The converse has 4 steps:
\begin{enumerate}
\item By \nameref{ax:Monot}, preference on the standard lotteries $L_t$ increases with $t$. \label{step:monotonicity}
\item By \nameref{ax:Continuity}, for any reward $r$ there is a unique $t\in[0,1]$ such that $r \sim L_t$, and we set $u(r)=t$.  \label{step:continuity}
\item Given a lottery $A$, each of its rewards $r$ is equivalent to $L_{u(r)}$. Hence  \nameref{ax:Subst}, and the way classical probabilities combine, imply $A$ is equivalent to $L_{EU(A)}$. \label{step:substitutability} 
\item By steps 1 and 3, preference on lotteries increases with their expected utilities.  
\end{enumerate}
\vspace{-\baselineskip}
\end{proof}

In decision under risk, a preference order satisfying the Von Neumann-Morgenstern axioms is also called \emph{rational}. Implicit in such label is the idea that an intelligent decision maker ought to follow the axioms. But empirical research shows that even  smart educated people often deviate from them, specially in cases with big rewards and small probabilities (e.g. in the Allais paradox). Many authors dismiss this, arguing that those agents, when confronted with their error, would recognize it and correct their decision. Others, however, question the validity of the axioms, in particular the \nameref{ax:Indep} one.

\subsection{Decision under uncertainty}\label{sec:Decision under uncertainty}

In problems of decision under uncertainty there are multiple possible outcomes, but their probabilities are unknown.
Instead of lotteries, problems are usually framed in terms of a preference order on \emph{acts} which, depending on possible \emph{states of the world}, will lead to \emph{payoffs} (consequences).

Some decision strategies 
(\emph{maximax}, \emph{maximin}, \emph{minimax regret}, etc.) focus on best or worst cases. 
These are useful in some situations, but can lead to absurd results in others, as they disregard important data, like non-extremal payoffs, or the low likelihoods of some states of the world.

Another strategy uses subjective probabilities, estimates by the agent of the likelihoods of states of the world. But the result can be very sensitive to the estimates used, especially if there are unlikely states with huge payoffs. Despite this, L.\,Savage \cite{Savage1972} proved, in 1954, that any preference order satisfying certain axioms can be induced, via the Principle of Maximization of Expected Utility, by some utility function and subjective probabilities. 

Let $\mathcal{S}$ be the set of states of the world. Not all details of a state are relevant, so we consider \emph{events}, subsets of states with some common characteristic. For example, the event $E=$``it rains tomorrow'' consists of all states in which this happens. Events can be partitioned into smaller ones, e.g. $E$ could have subevents like ``it rains tomorrow and stock prices go up'' and ``it rains tomorrow and the result of a die is 3''. Events can be combined through intersections, unions and complements, corresponding to the logical operators AND, OR, and NOT.

An\emph{ act} is a function $f:\mathcal{S}\rightarrow\mathcal{P}$, where $\mathcal{P}$ is the set of payoffs, so that $f(s)=x$ means $x$ is the payoff resulting from act $f$ if the state of world turns out to be $s$. Each payoff $x$ is identified with a \emph{single payoff act}, which results $x$ for all states.
Given acts $f$ and $g$, and an event $E$, the \emph{compound act} $[E,f,g]$ is 
\[ [E,f,g](s)=\begin{cases}
f(s)\ \text{ if } s\in E, \\
g(s)\ \text{ if } s\notin E.
\end{cases} \] 

Savage adopts, for a preference order $\succ$ on acts, the following axioms\footnote{The order and names of the axioms vary in the literature. We adapted that of \cite{Karni2014}.}:

\begin{Axiom}[S1 (Order)]\label{ax:S1}
$\succ$ is complete and transitive.
\end{Axiom}
\begin{Axiom}[S2 (Sure-Thing Principle)]\label{ax:S2}
For any event $E$ and acts $f,g,h$ and $k$, we have $[E,f,h]\succcurlyeq [E,g,h] \Leftrightarrow [E,f,k]\succcurlyeq [E,g,k].$ 
\end{Axiom}
This is similar to \nameref{ax:Indep}: preference between acts depends only on the events where they differ. It allows us to define, for each event $E$, a conditional preference order $\succ_{E}$ by 
\[ f\succ_{E}g \text{ \ if and only if \ } [E,f,h]\succ[E,g,h] \text{ \ for all }h,\]
meaning ``$f$ is preferable to $g$ given $E$'', i.e. if the agent assumes $E$ will happen he will prefer $f$ to $g$. 

An event $E$ is \emph{null} if $f\sim_{E}g$ for all $f$ and $g$, i.e. the agent does not care about acts on $E$ (we will find the reason is he does not believe $E$ can happen). Nullity has the following properties, for events $E$ and $F$:
\begin{ThmName}[Classical Null Subevent]\label{pr:Classical Null Subevent}
If $E$ is null, and $F\subset E$, then $F$ is null.
\end{ThmName}
\begin{ThmName}[Classical Null Disjunction]\label{pr:Classical Null Disjunction}
If $E$ and $F$ are null then $E\cup F$ is null.
\end{ThmName}
They result from properties of sets and functions, like the fact that a function can be redefined in a subset, without affecting the complement. We prove the last one, for comparison with the quantum case.
\begin{proof}
Without loss of generality, we can assume $E\cap F=\emptyset$. Given acts $f,g$ and $h$, define a new act $k$ by
\[ k(s)=\begin{cases}
g(s)\ \text{ if } s\in E, \\
f(s)\ \text{ if } s\in F, \\
h(s)\ \text{ if } s\notin E\cup F.
\end{cases} \]
As $E$ and $F$ are null, $[E\cup F,f,h]\sim k \sim [E\cup F,g,h]$. 
\end{proof}

\begin{Axiom}[S3 (Ordinal Event Independence)]\label{ax:S3}
For any payoffs $x$ and $y$, and any non null event $E$, we have $x \succ_{E} y \,\Leftrightarrow\, x\succ y$. 
\end{Axiom}

Hence the agent's preference order on payoffs is independent of any given event. This is not as banal as it may seem: receiving an umbrella might be preferable than sunglasses in the event of rain, but the preference might be reversed in a sunny day. So this axiom demands a careful consideration of what we are to call acts and payoffs. In this example, possession of umbrellas or sunglasses ought to be acts, with getting wet or having protected eyes being payoffs.

\begin{Axiom}[S4 (Comparative Probability)]\label{ax:S4}
Let $x,y,z$ and $w$ be payoffs, and $E,F$ be events. If $x\succ y$ and $z \succ w$, then $[E,x,y]\succ[F,x,y] \,\Leftrightarrow\, [E,z,w]\succ[F,z,w]$.
\end{Axiom}

The intuitive idea is as follows. $[E,x,y]$ gives a better payoff in case of $E$ than otherwise. $[F,x,y]$ gives the same payoffs, but the event giving the better payoff is $F$ instead of $E$. By \nameref{ax:S3} the agent only cares about the payoff and not how he got it, so the only reason for preferring $[E,x,y]$ to $[F,x,y]$ is that he thinks $E$ is more likely than $F$. So the same preference should hold for any other pair $z$, $w$ of better/worse payoffs.

This induces an order on events by
\[ E\succ F \text{ \ if and only if \ } [E,x,y]\succ[F,x,y] \text{ whenever } x\succ y. \]
It means the agent thinks $E$ is more likely than $F$, as he prefers the better payoff be given in case of $E$ than in case of $F$. 
Savage proves this order is a \emph{qualitative probability} on events, i.e. it satisfies
\begin{itemize}
\item[(i)] $\succ$ is complete and transitive;
\item[(ii)] $E\succcurlyeq\emptyset$ for any $E\subset\mathcal{S}$; 
\item[(iii)] $\mathcal{S}\succ\emptyset$; 
\item[(iv)] for any $E,E',F\subset\mathcal{S}$ with $E\cap F=E'\cap F=\emptyset$, $E\succ E' \Leftrightarrow E\cup F \succ E'\cup F$. 
\end{itemize}  

Actually, for this to work we need at least one pair of better/worse payoffs, given by the next axiom.

\begin{Axiom}[S5 (Nondegeneracy)]\label{ax:S5}
There are payoffs $x$, $y$ such that $x \succ y$.
\end{Axiom}
If this axiom is violated, we have a trivial decision problem. In such case, most of \nameref{th:Savage} is trivially true, but we do not obtain a unique subjective probability measure (in fact, any probability measure would work).

The next axiom allows us to turn the qualitative probability into a quantitative one.

\begin{Axiom}[S6 (Small-Event Continuity)]\label{ax:S6}
If $f\succ g$ then for any payoff $x$ there is a finite partition $\mathcal{S}=\mathbin{\mathaccent\cdot\cup}_{i=1}^n E_i$ such that $[E_i,x,f]\succ g$ and $f\succ[E_i,x,g]$ for every $i$.
\end{Axiom}

This is similar to the \nameref{ax:Arch}. It means the set of states of the world can be partitioned in events $E_i$ deemed so unlikely that changing the payoff on one of them is not enough to alter preferences. And no payoff is infinitely good or bad, or it would alter preferences for any non null $E_i$. The partition could be given, for example, by the results of roulettes with arbitrarily large numbers of slots, or by throwing dies any number of times.

Whenever $F\succ E$, the axiom implies there is a finite partition $\mathcal{S}=\mathbin{\mathaccent\cdot\cup}_{i=1}^n E_i$ such that $F\succ E\cup E_i$ for any $i$. 
Some $E_i$'s may be deemed more likely than others, but, after some technical work, Savage obtains partitions whose events the agent considers as equally likely as necessary.

Partitioning $\mathcal{S}$ into $n$ events considered equally likely, we attribute to each a subjective probability $\frac{1}{n}$. If, for a large $n$, the minimum number of these pieces needed to cover some event $E$ is $m$, its subjective probability should be close to $\frac{m}{n}$. The exact value is defined via a limiting process.
In particular, an event is null if, and only if, its subjective probability is $0$.

One last axiom is needed for situations with infinitely many payoffs. It requires that if $f$ is preferred, given $E$, to all payoffs $g$ can give in such event, then $f$ is preferred to $g$ given $E$.

\begin{Axiom}[S7 (Dominance)]\label{ax:S7}
If $f \succ_E g(s)$ for all $s\in E$, then $f \succ_E g$ (and likewise for $\prec_E$).
\end{Axiom}

Once we have the agent's subjective probabilities, the problem becomes one of decision under risk. 
Savage's result is similar to the \nameref{th:VNM}, but with subjective probabilities.
\begin{ThmName}[Savage's Theorem]\label{th:Savage}
A preference relation $\succ$ on acts satisfies axioms S1$-$S7 if, and only if, there is an unique, nonatomic\footnote{This means any non null event can be partitioned into two non null subevents.}, finitely additive probability measure $\pi$ on $\mathcal{S}$, and a bounded utility function $u$ on $\mathcal{P}$, such that $\succ$ is represented by $u$ via the Principle of Maximization of Expected Utility (with expected utilities defined in terms of the subjective probability $\pi$). Moreover, $u$ is unique up to positive affine transformations, and an event $E$ is null if, and only if, $\pi(E)=0$.
\end{ThmName}

Savage's axioms are usually taken as mandates of rationality for decision under uncertainty, even if they may appear less natural than those in the previous cases, and even though intelligent people often violate them (e.g. in the Ellsberg paradox).

\bibliographystyle{amsalpha}  
\bibliography{../Bibliografia_Wallace}

\end{document}